\documentclass[a4paper,11pt]{article}
\pdfoutput=1 % if your are submitting a pdflatex (i.e. if you have
\usepackage{jheppub} % for details on the use of the package, please
\usepackage[T1]{fontenc} % if needed
\usepackage{multirow}%表格合并
\usepackage[normalem]{ulem}
\usepackage{graphicx}
\usepackage{amsmath,amssymb}
\usepackage{dcolumn}
\usepackage{bm}
\usepackage{color}

\newcommand{\f}{\frac}
\newcommand{\lt}{\left}
\newcommand{\m}{m_{\rm P}}
\newcommand{\n}{\nonumber}
\newcommand{\p}{\partial}
\newcommand{\rt}{\right}
\newcommand{\dd}{{\rm d}}
\newcommand{\bt}{\beta}
\newcommand{\dt}{\delta}

\newcommand{\ve}{\varepsilon}
\newcommand{\sg}{\sigma}

\newcommand{\pb}{{\rm PBH}}
\newcommand{\cR}{{\cal R}}
\newcommand{\cP}{{\cal P}}
\newcommand{\la}{\langle}
\newcommand{\ra}{\rangle}
\newcommand{\arxgr}[1]{\href{http://arxiv.org/abs/#1}{{\ttfamily arXiv:#1[gr-qc]}}}
\newcommand{\arxph}[1]{\href{http://arxiv.org/abs/#1}{{\ttfamily arXiv:#1[hep-ph]}}}
\newcommand{\arxth}[1]{\href{http://arxiv.org/abs/#1}{{\ttfamily arXiv:#1[hep-th]}}}
\newcommand{\arxas}[1]{\href{http://arxiv.org/abs/#1}{{\ttfamily arXiv:#1[astro-ph]}}}
\newcommand{\Arxgr}[1]{\href{http://arxiv.org/abs/gr-qc/#1}{{\ttfamily arXiv:#1[gr-qc]}}}
\newcommand{\Arxth}[1]{\href{http://arxiv.org/abs/hep-th/#1}{{\ttfamily arXiv:#1[hep-th]}}}
\newcommand{\Arxas}[1]{\href{http://arxiv.org/abs/astro-ph/#1}{{\ttfamily arXiv:#1[astro-ph]}}}
\newcommand{\https}[1]{\href{https://#1}{{\ttfamily https://#1}}}

\title{\boldmath Primordial black holes from the perturbations in the inflaton potential in peak theory}
\author[a]{Qing Wang}
\author[a]{Yi-Chen Liu}
\author[a]{Bing-Yu Su}
\author[a,1]{and Nan Li \note{Corresponding author.}}
\affiliation[a]{Department of Physics, College of Sciences, Northeastern University \\ No. 3-11, Wenhua Road, Shenyang, 110819, China}
\emailAdd{linan@mail.neu.edu.cn}
\abstract{The primordial black hole (PBH) is an effective candidate for dark matter. In this work, the PBH abundance $f$ is calculated in peak theory, with one or two perturbations in the inflaton potential. We construct an antisymmetric perturbation that can create a perfect plateau in the inflaton potential, leading inflation to the ultra-slow-roll stage. During this stage, the power spectrum of primordial curvature perturbation is remarkably enhanced on small scales, generating abundant PBHs. The PBH abundance $f\sim 0.1$ can be achieved in one or two typical mass windows at $10^{-17}M_\odot$, $10^{-13}M_\odot$, and $30M_\odot$, without spoiling the nearly scale-invariant power spectrum on large scales. For comparison, $f$ is calculated in two approximate methods of peak theory (with different spectral moments) and also in the Press--Schechter theory. It is found that the Press--Schechter theory systematically underestimates $f$ by two or three orders of magnitude compared with peak theory.}

\begin{document}
\maketitle
\flushbottom

\section{Introduction} \label{sec:intro}

Primordial black holes (PBHs) have been receiving increasing research interest in recent years, especially after the discovery of the long-sought gravitational waves (GWs) from the merger of binary black holes \cite{LIGO}. The influences of PBHs in cosmology are manifold. For instance, the PBHs can be the sources of GWs \cite{gw}, and their Hawking radiation can also change the background intensities of various cosmic rays \cite{zs}. Moreover, they can be utilized as a probe in the very early Universe to constrain the inflation models \cite{Khlopov:2008qy, Green:2020jor}. Most importantly, the PBH is a natural candidate for dark matter (DM) (e.g., as the seeds of super-massive black holes in the centers of galaxies) \cite{Belotsky:2014kca, dm}.

PBHs can be formed in many different ways in the radiation-dominated era of the early Universe, such as various phase transitions \cite{zs, Green:2020jor, dm}. Nevertheless, the simplest picture is the direct collapse of the radiation field, when its density fluctuation exceeds some certain threshold \cite{Zel, hkkk, carrhkkk, Carr:1975qj}. Different from astrophysical black holes, PBHs have a vast mass range spanning over 40 orders of magnitude, from the Planck mass to super-massive scale ($10^{-5}\sim 10^{40}$ g), without the lower bound around $3M_\odot$ or the mass gap around $50\sim 120M_\odot$ ($M_\odot=1.989\times10^{33}~{\rm g}$ denotes the solar mass). Because of the Hawking radiation, PBHs with $M<10^{15}$ g have already evaporated, and PBHs with $M>10^{15}$ g can still exist today, acting as a stable and pressureless candidate for DM.

The PBH abundance $f$ is defined as its proportion in the DM density at present, and the experimental constraints on PBHs refer to the upper bounds of $f$ in different mass ranges. If $f\sim 0.1$, PBHs can be considered as an effective candidate for DM; if $f\ll 10^{-3}$, its possibility as DM can be safely excluded from the relevant mass range. According to various constraints from the evaporation, lensing, dynamics, accretion, and GW experiments, there still remain three open mass windows: the asteroid mass range ($10^{-17}\sim 10^{-16}M_\odot$), the sub-lunar mass range ($10^{-13}\sim 10^{-9}M_\odot$), and the intermediate mass range ($10\sim 10^3M_\odot$) \cite{dm}. (For more recent and complete constraints on the PBHs, see \cite{bound}.)

Generally speaking, in order to achieve the PBH abundance $f\sim 0.1$, the radiation field must possess a sufficiently large density fluctuation, with the power spectrum of primordial curvature perturbation $\cP_\cR$ enhanced up to at least $10^{-2}$ on small scales. However, such a large magnitude cannot be obtained in the usual single-field slow-roll (SR) inflation models, which generate a nearly scale-invariant power spectrum only on large scales [e.g., $\cP_\cR$ is just around $10^{-9}$ on the pivot scale $k_\ast$ of the cosmic microwave background (CMB) \cite{Planck}]. Consequently, the SR conditions should be violated on small scales, and this violation can be realized by the so-called ultra-slow-roll (USR) stage in the inflationary era \cite{Ivanov:1994, GarciaBellido:1996qt, Leach:2001zf, Kinney:2005vj, Chongchitnan:2006wx, Martin:2012pe, Clesse:2015wea, Garcia-Bellido:2017mdw, Kannike:2017bxn, Germani:2017bcs, Motohashi:2017kbs, Dimopoulos:2017ged, Ezquiaga:2017fvi, Ballesteros:2017fsr, Cicoli:2018asa, Ozsoy:2018flq, Gong:2017qlj, Biagetti:2018pjj, Gao:2018pvq, Germani:2018jgr, Dalianis:2018frf, Cruces:2018cvq, Passaglia:2018ixg, Ballesteros:2018wlw, Cheng:2018qof, Byrnes:2018txb, Vallejo-Pena:2019lfo, Pattison:2019hef, Fu:2019ttf, Carrilho:2019oqg, Bhaumik:2019tvl, Motohashi:2019rhu, Mahbub:2019uhl, liu, Lin:2020goi, Mishra:2019pzq, Cai:2019bmk, Ballesteros:2020sre, Kefala:2020xsx, Figueroa:2020jkf, Ozsoy:2020kat, Tasinato:2020vdk, Ragavendra:2020sop, Pattison:2021oen, Choi:2021yxz, Solbi:2021wbo, Inomata:2021uqj, Kawai:2021bye, Zheng:2021vda, Zhang:2021vak, Liu:2021rgq, Heydari:2021gea, Kawai:2021edk, Artigas:2021zdk, Heydari:2021qsr}.

If there is a plateau or a saddle point in the single- or multi-field inflaton potential $V$, the inflaton will roll down its potential extremely slowly, resulting in a remarkable enhancement of $\cP_\cR$ and a considerable PBH abundance simultaneously. The USR conditions can be realized in many ways. For example, one may consider the (near-)inflection points in $V$, but the specific form of $V$ must be designed very carefully, so that the USR inflation increases $\cP_\cR$ only on small scales but does not spoil the CMB physics on large scales. Instead, in this paper, we take another approach and investigate the USR inflation and the PBH abundance by imposing one or two perturbations $\dt V$ on the background potential $V$. By this means, the USR inflation can be studied on small and large scales separately, without the troublesome interference in between \cite{Ozsoy:2018flq, Mishra:2019pzq, Ozsoy:2020kat, Zheng:2021vda, Zhang:2021vak}. The essential difference between our work and the previous literature is a new construction of the perturbations. Usually, these perturbations were chosen to be symmetric (e.g., Gaussian) in the USR region, but $V$ itself must not be symmetric there---otherwise, the inflaton cannot roll down at all. Therefore, we now choose an antisymmetric perturbation that can be imposed on $V$ more smoothly on both sides of the USR region. As a result, there can be a plateau flat enough in $V$, and the USR stage can be maintained for a sufficiently long time, generating the PBHs with desirable masses and abundances.

In this paper, we numerically calculate the power spectrum $\cP_\cR$ and the PBH abundance $f$ in the single-field inflation models, with one or two perturbations in the inflaton potential, as the analytical SR approximations can no longer be trusted in the USR stage. We demand the PBH abundance $f\sim 0.1$ in one or two typical mass windows at $10^{-17}M_\odot$, $10^{-13}M_\odot$, and $30M_\odot$. Here, we should mention that $f$ is affected by many factors, such as the choice of window function, the threshold of density contrast, and the method to calculate the PBH mass fraction. Below, we focus on the last aspect, calculate the PBH mass fraction and abundance in peak theory \cite{Doroshkevich, peak, Green:2004wb, Young:2014ana, Yoo:2018kvb, Yoo:2019pma, Germani:2019zez, Young:2020xmk, Wu:2020ilx, Yoo:2020dkz, Mahbub:2021qeo}, and compare the results with those obtained from the Press--Schechter (PS) theory \cite{PS}.

The peak theory of Gaussian random fields was initiated in Ref. \cite{peak}. Furthermore, in Ref. \cite{Green:2004wb}, Green, Liddle, Malik, and Sasaki (GLMS) calculated the PBH abundance via a simplified peak theory and arrived at the results in reasonable agreement with the PS theory, but only for the power spectrum in the power-law form. Their results were later extended to the scale-invariant power spectrum and the power spectrum with a running spectral index \cite{Young:2014ana}, and the PBH abundances obtained from peak and PS theories were found to be different but still similar, within a factor of ${\cal O}(10)$. The peak theory with local non-Gaussianity was discussed in Ref. \cite{Yoo:2019pma}, the point-peak approximation was presented in Ref. \cite{Wu:2020ilx}, and the optimized peak theory was also explored in Ref. \cite{Mahbub:2021qeo}. In general, the PBH abundances obtained in various peak theories are always larger than that from the PS theory, but the detailed ratio depends on the specific profile of density fluctuations \cite{Yoo:2018kvb, Yoo:2020dkz}.

The basic idea of peak theory is to calculate the number density of peaks where compact objects like galaxies and PBHs are expected to form, if the density contrast $\dt$ exceeds some threshold $\dt_{\rm c}$. As the peaks correspond to the local maxima of $\dt$, peak theory deals with a multi-dimensional joint probability distribution function (PDF) of $\dt$ and its first- and second-order spatial derivatives. In order to obtain the one-dimensional conditional PDF of $\dt$, we must perform the dimensional reduction on the multi-dimensional PDF by a series of integrations. The general peak theory is very complicated, and the approximations at different levels have been developed for various practical aims, especially for the large density contrast in the PBH formation. In these approximations, we are confronted with the spectral moments $\sg_i$ of density contrast at different orders. In Sec. \ref{sec:peak}, we refer the high-peak approximation to the one including $\sg_0$, $\sg_1$, and $\sg_2$ and the GLMS approximation to the one including $\sg_0$ and $\sg_1$. As a comparison, the simplest PS theory involves only $\sg_0$ and can be viewed as the limit case of the general peak theory. All these three cases will be discussed and compared in detail below.

This paper is organized as follows. In Sec. \ref{sec:pbh}, we study the power spectrum and calculate the PBH mass and abundance. In Sec. \ref{sec:peak}, we review the general peak theory and the relevant high-peak and GLMS approximations. The PS theory is also discussed for comparison. Then, in Sects. \ref{sec:one} and \ref{sec:two}, we study the power spectra and PBH abundances in the GLMS approximation, with one or two perturbations in the inflaton potential, respectively. Finally, a general comparison of the PBH abundances in peak and PS theories is presented in Sec. \ref{sec:bj}. We conclude in Sec. \ref{sec:con}. We work in the natural system of units and set $c=\hbar=k_{\rm B}=1$.

\section{PBH abundance} \label{sec:pbh}

In this section, we discuss the power spectrum of primordial curvature perturbation, explain the significant difference between the SR and USR inflations, and calculate the PBH mass and abundance in detail.

\subsection{Basic equations} \label{sec:basic}

We start from the single-field inflation model, in which inflation is supported by a scalar inflaton field minimally coupled to gravity, with the action reading
\begin{align}
S=\int\dd^4 x\,\sqrt{-g}\lt[\f{m_{\rm P}^2}{2}R-\frac{1}{2}\p_\mu\phi\p^\mu\phi-V(\phi)\rt], \n
\end{align}
where $\phi$ is the inflaton field, $V(\phi)$ is its potential, $R$ is the Ricci scalar, and $m_{\rm P}=1/\sqrt{8\pi G}$ is the Planck mass. In the inflationary era, it is more convenient to measure the cosmic expansion in terms of the number of $e$-folds $N$ instead of the cosmic time $t$, which is defined as $\dd N=H(t)\,\dd t=\dd\ln a$. Here, $a=e^N$ is the scale factor, and $H=\dot{a}/a$ is the Hubble expansion rate. In order to solve the flatness and horizon problems in the standard Big Bang cosmology, the Universe must experience a quasi-de Sitter expansion, with $N$ being around 70.

Two useful parameters can be introduced to characterize the motion of the inflaton field,
\begin{align}
\ve=-\frac{\dot H}{H^2}, \quad \eta=-\frac{\ddot \phi}{H\dot\phi}. \n
\end{align}
Usually, they are named as the SR parameters, as in the SR inflation, $\ve$ and $|\eta|$ are both much smaller than 1. However, they can also be applied to other circumstances like the USR stage, during which they change drastically, and this is of essential importance in calculating the PBH abundance.

In terms of $\ve$, $\eta$, and $N$, in the homogeneous and isotropic Universe, the Klein--Gordon equation for the inflaton field $\phi$ is
\begin{align}
\phi_{,NN}+(3-\ve)\phi_{,N}+\frac{1}{H^2}V_{,\phi}=0, \label{kgphi}
\end{align}
and the Friedmann equation for the cosmic expansion is
\begin{align}
H^2=\frac{V}{(3-\ve)m^2_{\rm P}}. \label{Fried}
\end{align}
Furthermore, we study the perturbations on the background space-time, with the perturbed metric reading $\dd s^2=-(1+2\Phi)\,\dd t^2+a^2(t)(1-2\Phi)\dt_{ij}\,\dd x^i\dd x^j$. Here, we neglect the anisotropic stress and focus only on the scalar perturbation $\Phi$, as the vector and tensor perturbations are irrelevant to the production of PBHs. A more useful gauge-invariant scalar perturbation is the primordial curvature perturbation $\cR$, which can be used to calculate the PBH abundance,
\begin{align}
{\cal R}=\Phi+\frac{H}{\dot{\phi}}\dt\phi. \n
\end{align}
The equation of motion for $\cR$ in the Fourier space is the Mukhanov--Sasaki equation,
\begin{align}
{\cal R}_{k,NN}+(3+\ve-2\eta){\cal R}_{k,N}+\frac{k^2}{H^2e^{2N}}{\cal R}_k=0. \label{MS}
\end{align}
Moreover, $\ve$ and $\eta$ can also be expressed in terms of $N$,
\begin{align}
\ve=\frac{\phi_{,N}^2}{2\m^2}, \quad \eta=\frac{\phi_{,N}^2}{2m_{\rm P}^2}-\frac{\phi_{,NN}}{\phi_{,N}}. \label{eta}
\end{align}
Altogether, we will numerically solve Eqs. (\ref{kgphi})--(\ref{eta}) to calculate the power spectrum and the PBH abundance in the following sections.

\subsection{Power spectrum} \label{sec:power}

The statistical properties of Gaussian random fields are completely encoded in the two-point correlation function (i.e., the power spectrum in the Fourier space). The PBH abundance can be calculated via the dimensionless power spectrum of primordial density contrast $\cP_{\dt}(k)$. In the radiation-dominated era, $\cP_{\dt}(k)$ can be further related to the dimensionless power spectrum of primordial curvature perturbation $\cP_{\cR}(k)$ \cite{Green:2004wb},
\begin{align}
\cP_{\dt}(k)=\f{16}{81}\lt(\f{k}{aH}\rt)^4\cP_{\cR}(k), \n
\end{align}
where
\begin{align}
\lt.{\cal P}_{\cal R}(k)=\f{k^3}{2\pi^2}|{\cal R}_{k}|^2\rt|_{k\ll aH}.\n
\end{align}

In the SR inflation, ${\cal P}_{\cal R}(k)$ is nearly scale invariant,
\begin{align}
{\cal P}_{\cal R}(k)=\f{1}{8\pi^2\ve}\lt(\f{H}{\m}\rt)^2, \label{PRsr}
\end{align}
and can be usually written in the power-law form as ${\cal P}_{\cal R}(k)=A_{\rm s}({k}/{k_\ast})^{n_{\rm s}-1}$, where $n_{\rm s}$ is the scalar spectral index and $A_{\rm s}$ is the amplitude, with the central values being $n_{\rm s}=0.965$ and $A_{\rm s}=2.10\times 10^{-9}$ on the CMB pivot scale $k_*=0.05~{\rm Mpc}^{-1}$ \cite{Planck}. However, on small scales where PBHs are expected, the constraints on ${\cal P}_{\cal R}(k)$ are still rather loose \cite{zs, Green:2020jor, dm} (the CMB observations cover only about 10 $e$-folds on large scales). In particular, in the USR region, both numerical and analytical investigations indicate $n_{\rm s}-1 \approx 4$ in the growing stage of ${\cal P}_{\cal R}(k)$ \cite{Byrnes:2018txb, liu}.

Now, we first need to smooth the density contrast $\dt$ on a large scale, usually taken as $R=1/(aH)$, in order to avoid the non-differentiability and the divergence in the large-$k$ limit of the radiation field. This smoothing procedure can be realized by a convolution of $\dt$ as $\dt({\bf x},R)=\int\dd^3x'\, W({\bf x}-{\bf x}',R)\dt({\bf x}')$, with $W({\bf x},R)$ being the window function. Below, we choose Gaussian window function, with its Fourier transform being $\widetilde{W}(k,R)=e^{-k^2R^2/2}$. Hence, the window function in real space is $W({\bf x},R)=e^{-{x^2}/({2R^2})}/V(R)$, and the volume $V(R)=(\sqrt{2\pi}R)^3$ is the normalization factor.

The variance of the smoothed density contrast on the scale $R$ can be calculated as
\begin{align}
\sigma_{\dt}^2(R)=\la\dt^2({\bf x},R)\ra=\int_{0}^{\infty}\f{\dd k}{k}\,\widetilde{W}^2(k,R)\cP_{\dt}(k), \n
\end{align}
where $\la\cdots\ra$ denotes the ensemble average, and we have used the fact $\la\dt({\bf x},R)\ra=0$ for a Gaussian random field. Moreover, the homogeneity and isotropy of the background Universe guarantees that $\sigma_{\dt}^2(R)$ is independent of a special position ${\bf x}$. Similarly, the $i$-th spectral moment of the smoothed density contrast is defined as
\begin{align}
\sigma_{i}^2(R)&=\int_{0}^{\infty}\f{\dd k}{k}\,k^{2i}\widetilde{W}^2(k,R)\cP_{\dt}(k)
=\f{16}{81}\int_{0}^{\infty}\f{\dd k}{k}\,k^{2i}\widetilde{W}^2(k,R)(kR)^4\cP_{\cR}(k), \n
\end{align}
where $i=0,1,2,\cdots$, and $\sigma_0=\sigma_\dt$ naturally.

For the convenience of expression in Sec. \ref{sec:peak}, two new important factors in peak theory are introduced here. The first one is the relative density contrast,
\begin{align}
\nu=\f{\dt}{\sigma_{\dt}}, \n
\end{align}
and its corresponding threshold is $\nu_{\rm c}=\dt_{\rm c}/\sigma_{\dt}$. The specific value of $\dt_{\rm c}$ depends on the equation of state of the cosmic medium and many other ingredients \cite{Niemeyer:1999ak, Musco:2004ak, Musco:2008hv, Musco:2012au, 414, Nakama:2013ica, Musco:2018rwt, Escriva:2019nsa, Escriva:2019phb, Escriva:2020tak, Musco:2020jjb} and is the most influential factor in calculating the PBH abundance. In this paper, we follow Ref. \cite{414} and set $\dt_{\rm c}=0.414$. In Sec. \ref{sec:bj}, it will be shown that $\nu_{\rm c}$ should be around ${\cal O}(10)$, so as to achieve abundant PBHs, and $\nu_{\rm c}$ is not a constant, as $\sg_\dt$ depends on the smoothing scale $R$. Furthermore, the second factor is defined as
\begin{align}
\gamma=\f{\sigma_{1}^{2}}{\sigma_{\dt}\sigma_{2}}. \n
\end{align}
The $\gamma$ factor encodes the specific information of the profile of density contrast, and it is easy to find $0<\gamma<1$ by definition. Also, in Sec. \ref{sec:bj}, it will be shown that $\gamma\approx 1$ for the PBH formation.

\subsection{PBH mass and abundance} \label{sec:mass}

In the Carr--Hawking collapse model \cite{carrhkkk}, the PBH mass $M$ can be related to the horizon mass at the time of its formation,
\begin{align}
M=\kappa M_{\rm H}=\f{\kappa}{2GH}, \n
\end{align}
where $M_{\rm H}=1/(2GH)$ is the horizon mass, and $\kappa$ is the efficiency of collapse. In the radiation-dominated era, $H=1/(2t)$, so $M=\kappa{t}/{G}$. From the conservation of entropy in the adiabatic cosmic expansion, we have \cite{zs}
\begin{align}
\f{M}{M_{\odot}}=1.13\times10^{15}\lt(\f{\kappa}{0.2}\rt)\lt(\f{g_{\ast}}{106.75}\rt)^{-1/6}\lt(\f{k_{\ast}}{k_{\pb}}\rt)^{2}, \label{M}
\end{align}
where $g_{\ast}$ is the effective number of relativistic degrees of freedom of energy density, and $k_{\pb}=1/R$ is the wave number of the PBH that exits the horizon. Below, we follow Ref. \cite{Carr:1975qj} and choose $\kappa=0.2$ and $g_{\ast}=106.75$. From Eq. (\ref{M}), all spectral moments $\sg_i(R)$ can be reexpressed in terms of the PBH mass as $\sg_i(M)$.

Furthermore, the PBH mass fraction $\bt(M)$ at the time of its formation is defined as
\begin{align}
\lt.\bt(M)=\f{\rho_\pb(M)}{\rho_{\rm R}}\rt|_{\rm formation}. \n
\end{align}
For the massive PBHs not evaporated yet (ignoring their radiation, accretion, and merger), their abundance at present is defined as
\begin{align}
\lt.f(M)=\f{\rho_{\pb}(M)}{\rho_{\rm DM}}\rt|_{\rm today}. \n
\end{align}
Naturally, $f(M)$ is proportional to $\bt(M)$ \cite{zs},
\begin{align}
f(M)=1.68\times10^{8}\lt(\f{M}{M_\odot}\rt)^{-1/2}\beta(M). \label{zheteng}
\end{align}

Here, we should stress that the PBH mass $M$ has the same expression in peak and PS theories, but the PBH mass fraction $\beta$ and abundance $f$ are different in various peak theories and the PS theory. All these issues will be systematically investigated in this paper.

\section{Peak theory and PS theory} \label{sec:peak}

In this section, we briefly review the peak theory of Gaussian random fields and present the general formula for the PBH mass fraction $\beta$. For comparison, the simplified results of $\beta$ are shown in the high-peak approximation, the GLMS approximation, and the PS theory, respectively.

\subsection{Peak theory} \label{sec:gpk}

The general peak theory of Gaussian random fields in Ref. \cite{peak} can also be applied to calculate the PBH abundance, with the peak value being the relative density contrast $\nu$. The number density of peaks is $n({\bf r})=\sum_p\dt_{\rm D}({\bf r}-{\bf r}_p)$, where $\dt_{\rm D}$ is the three-dimensional Dirac function, and ${\bf r}_p$ is the position where $\dt$ has a local maximum. This maximum condition further demands that the first-order derivative $\p_i\dt$ vanishes ($i=1,2,3$) and the second-order derivative $\p_i\p_j\dt$ is negative definite at ${\bf r}_p$ ($i,j=1,2,3$). Since $\dt$ is assumed to be Gaussian, $\p_i\dt$ and $\p_i\p_j\dt$ are also Gaussian accordingly. Altogether, we should deal with the ten-dimensional joint PDF $P(\{y_i\})$ of Gaussian variables (one for $\dt$, three for $\p_i\dt$, and six for $\p_i\p_j\dt$ due to its symmetry),
\begin{align}
P(\{y_i\})=\f{1}{\sqrt{(2\pi)^{10}\det{\cal M}}}\exp\Bigg(\f 12\sum_{ij}\Delta y_i{\cal M}^{-1}_{ij}\Delta y_j\Bigg), \n
\end{align}
where ${\cal M}$ is the covariance matrix, and $\Delta y_i=y_i-\la y_i\ra$, with $y_1=\dt$, $y_2=\p_1\dt$, $\cdots$, $y_5=\p_1\p_1\dt$, $\cdots$, and $y_{10}=\p_2\p_3\dt$.

From Ref. \cite{peak}, a series of dimensional reductions can finally lead the ten-dimensional joint PDF $P(\{y_i\})$ to the one-dimensional conditional PDF $P(\nu)$. By means of $P(\nu)$, the number density of peaks $n(\nu_{\rm c})$ with $\nu>\nu_{\rm c}$ can be written in a more convenient integral form of the one-dimensional differential number density ${\cal N}(\nu)$,
\begin{align}
n(\nu_{\rm c})=\int_{\nu_{\rm c}}^{\infty}{\cal N}(\nu)\,\dd\nu=\f{1}{(2\pi)^2R_\ast^3}
\int_{\nu_{\rm c}}^{\infty}G(\gamma,\nu)e^{-\nu^2/2}\,\dd\nu, \n
\end{align}
where $R_{\ast}=\sqrt{3}\sigma_{1}/\sigma_{2}$, and
\begin{align}
G(\gamma,\nu)=\int_{0}^{\infty}\f{f(x)}{\sqrt{2\pi(1-\gamma^2)}}\exp\lt[\f{-(x-\gamma\nu)^2}{2(1-\gamma^2)}\rt]\,\dd x, \label{G}
\end{align}
with
\begin{align}
f(x)&=\f{x^3-3x}{2}\lt[{\rm erf}\lt(\sqrt{\f{5}{2}}x\rt)+{\rm erf}\lt(\sqrt{\f{5}{8}}x\rt)\rt] + \n\\ &\quad\sqrt{\f{2}{5\pi}}\lt[\lt(\f{31x^2}{4}+\f{8}{5}\rt)e^{-{5x^2}/{8}}+\lt(\f{x^2}{2}-\f{8}{5}\rt)e^{-{5x^2}/{2}}\rt]. \n
\end{align}
The constraint condition that the peak position corresponds to the local maximum of density contrast is encoded in the $G(\gamma,\nu)$ function implicitly.

Finally, the PBH mass fraction at the time of its formation can be obtained in peak theory as
\begin{align}
\beta=n(\nu_{\rm c})V(R)=\f{1}{\sqrt{2\pi}}\lt(\f{R}{R_\ast}\rt)^3\int_{\nu_{\rm c}}^{\infty}G(\gamma,\nu)e^{-\nu^2/2}\,\dd \nu. \label{betapeak}
\end{align}
In general, there are the zeroth, first, and second spectral moments $\sg_\dt$, $\sg_1$, and $\sg_2$ in $\beta$ implicitly. Retaining these moments to different orders corresponds to different approximate peak theories and induces different formulae for $\beta$, to be explained explicitly below.

\subsection{High-peak approximation} \label{sec:hpk}

The high-peak approximation keeps all three spectral moments $\sg_\dt$, $\sg_1$, and $\sg_2$, and the approximation refers to the condition $\gamma\nu\gg 1$. As $0<\gamma<1$, we have $\nu\gg 1$ (i.e., $\dt\gg\sg_\dt$). This can always be guaranteed, as the PBH formation is a rather rare event.

In the limit of $\gamma\nu\gg 1$, the part of the integrand in Eq. (\ref{G}), $\f{1}{\sqrt{2\pi(1-\gamma^2)}} \exp\big[\f{-(x-\gamma\nu)^2}{2(1-\gamma^2)}\big]$, behaves as the Dirac function, so only the values around $x=\gamma\nu$ significantly contribute to the integral. When $x\gg 1$, the asymptotic expansion of $f(x)$ is $f(x)=x^3-3x$, so $G(\gamma,\nu)=(\gamma\nu)^3-3\gamma\nu$. Substituting $G(\gamma,\nu)$ into Eq. (\ref{betapeak}), we obtain the PBH mass fraction in the high-peak approximation,
\begin{align}
\beta_{\rm hp}=\f{1}{\sqrt{2\pi}}Q^3\lt(\nu_{\rm c}^2+2-\f{3}{\gamma^2}\rt) e^{-\nu_{\rm c}^2/2}, \label{hp}
\end{align}
where $Q={R\sigma_1}/(\sqrt{3}\sigma_\dt)$.

\subsection{GLMS approximation} \label{sec:glms}

A more frequently used peak theory is the GLMS approximation, which is a further step of the high-peak approximation. Besides the condition $\gamma\nu\gg1$, it also demands $\gamma=\sg_1^2/(\sg_\dt\sg_2)\approx 1$. This means that $\sg_2$ is not independent of $\sg_\dt$ and $\sg_1$, so only two spectral moments $\sg_\dt$ and $\sg_1$ are enough in the GLMS approximation.

In Refs. \cite{peak, Green:2004wb}, the $G(\gamma,\nu)$ function was obtained in the above approximations,
\begin{align}
G(\gamma,\nu)=\lt[\f{\la k^2(R)\ra}{3}\rt]^{3/2}R_\ast^3(\nu^3-3\nu), \n
\end{align}
where $\la k^2(R)\ra$ is a function of the smoothing scale $R$,
\begin{align}
\la k^2(R)\ra=\f{1}{\la\dt^2(R)\ra}\int_{0}^{\infty}\f{\dd k}{k}k^2\widetilde{W}^2(k,R)\cP_{\dt}(k)=\f{\sigma_1^2}{\sigma_\dt^2}. \n
\end{align}
Hence, we obtain $G(\nu)=\nu^3-3\nu$, so the $G(\gamma,\nu)$ function depends only on $\nu$ in the GLMS approximation. Substituting $G(\nu)$ into Eq. (\ref{betapeak}), we obtain the PBH mass fraction in the GLMS approximation,
\begin{align}
\beta_{\rm GLMS}=\f{1}{\sqrt{2\pi}}Q^3(\nu_{\rm c}^2-1) e^{-\nu_{\rm c}^2/2}. \label{GLMS}
\end{align}
It is straightforward to find that Eq. (\ref{hp}) is reduced to Eq. (\ref{GLMS}) when $\gamma\approx 1$.

\subsection{PS theory} \label{sec:ps}

In the PS theory, the $G(\gamma,\nu)$ function is simply taken as a constant, $G=(R_\ast/R)^3$, independent of $\gamma$ or $\nu$, so from Eq. (\ref{betapeak}), the PS theory involves only the zeroth spectral moment $\sg_0$ (i.e., the variance of density contrast $\sg_\dt$). Therefore, it is a non-physical dimensional reduction of the general peak theory and can be regarded as its limit case.

Substituting $G$ into Eq. (\ref{betapeak}), it is easy to find that the PS theory is equivalent to directly assuming the PDF $P(\nu)$ to be Gaussian, $P(\nu)=e^{-\nu^2/2}/\sqrt{2\pi}$. Considering $\nu_{\rm c}\gg 1$, we obtain the PBH mass fraction in the PS theory,
\begin{align}
\beta_{\rm PS}=\f12{\rm erfc}\lt(\f{\nu_{\rm c}}{\sqrt{2}}\rt)\approx \f{1}{\sqrt{2\pi}\nu_{\rm c}}e^{-\nu_{\rm c}^2/2}. \label{betaPS}
\end{align}

Summarizing and comparing the results in Eqs. (\ref{hp})--(\ref{betaPS}), we find that $\beta_{\rm GLMS}$ is the largest PBH mass fraction but is only slightly larger than $\beta_{\rm hp}$. Moreover, at the leading order of $\nu_{\rm c}$, we obtain
\begin{align}
\f{\beta_{\rm PS}}{\beta_{\rm hp}}\approx \f{\beta_{\rm PS}}{\beta_{\rm GLMS}}\approx\f{1}{(Q\nu_{\rm c})^3}. \label{ln}
\end{align}
In Sec. \ref{sec:bj}, it will be shown that the values of $1/(Q\nu_{\rm c})^3$ can be as small as $10^{-3}\sim 10^{-2}$. Consequently, there is a systematic bias between the PS theory and the high-peak or GLMS approximation, and the PS theory leads to a significant underestimation of the PBH abundance.

\section{PBHs from one perturbation in the inflaton potential} \label{sec:one}

From the discussions in Secs. \ref{sec:pbh} and \ref{sec:peak}, the procedure to calculate the PBH abundance can be summarized as
\begin{align}
\dt V(\phi)\to{\cal P}_{\cal R}(k)\to\sigma_i^2(M)\to\beta(M)\to f(M). \n
\end{align}
The discrepancy among the high-peak approximation, the GLMS approximation, and the PS theory lies in the intermediate steps ${\cal P}_{\cal R}(k)\to\sigma_i^2(M)\to\beta(M)$, and the first and last steps $\dt V(\phi)\to{\cal P}_{\cal R}(k)$ and $\beta(M)\to f(M)$ are the same under all circumstances. In this section, we first construct a suitable form for one perturbation $\dt V(\phi)$ in the inflaton potential and then achieve the PBH abundance $f\sim 0.1$ in the GLMS approximation in the three typical mass windows at $10^{-17}M_\odot$, $10^{-13}M_\odot$, and $30M_\odot$, respectively (the PBH abundances in the high-peak approximation and the PS theory will be discussed in Sec. \ref{sec:bj}). The difference between the power spectra $\cP_\cR(k)$ in the SR and USR inflations will also be illustrated.

In general, the specific form of the perturbation $\dt V(\phi)$ is not unique, as long as it can smooth the background inflaton potential $V_{\rm b}(\phi)$ at some position $\phi_0$. In this way, a plateau appears around $\phi_0$, leading inflation to the USR stage and thus enhancing the power spectrum and the PBH abundance. Previously, some attempts (e.g., Gaussian perturbation) were considered in Refs. \cite{Ozsoy:2018flq, Mishra:2019pzq, Ozsoy:2020kat, Zheng:2021vda, Zhang:2021vak}. Gaussian $\dt V(\phi)$ is symmetric in the USR region, but $V_{\rm b}(\phi)$ itself cannot be symmetric and must have a certain slope, if the inflaton can roll down there. Hence, $\dt V(\phi)$ and $V_{\rm b}(\phi)$ cannot be connected very smoothly on both sides of $\phi_0$. As a result, the inflaton either still rolls down $V_{\rm b}(\phi)$ rapidly after the USR stage or even cannot surmount the perturbation and stops in the USR region, resulting in eternal inflation.

Therefore, in this paper, we suggest a new antisymmetric form of the perturbation as
\begin{align}
\dt V(\phi)=-A(\phi-\phi_0)F\lt(\frac{\phi-\phi_0}{\sqrt{2}\sg}\rt), \label{F}
\end{align}
where $F$ is an even function for the argument and satisfies $\lim\limits_{x\to\infty}xF(x)=0$. There are three parameters in our model: $A$, $\phi_0$, and $\sigma$, characterizing the slope, position, and width of $\dt V(\phi)$, respectively. Thus, if $A$ is very close to $V_{\rm b,\phi}(\phi_0)$, a perfect plateau can be created around $\phi_0$. In this way, the inflaton can pass the perturbation definitely, and the USR stage can be maintained for a sufficiently long time, generating abundant PBHs. In Secs. \ref{sec:one} and \ref{sec:two}, we will choose $F$ to be Lorentzian and Gaussian functions, in order to verify the applicability and universality of our model.

\begin{figure}[htb]
\centering \includegraphics[width=0.65\linewidth]{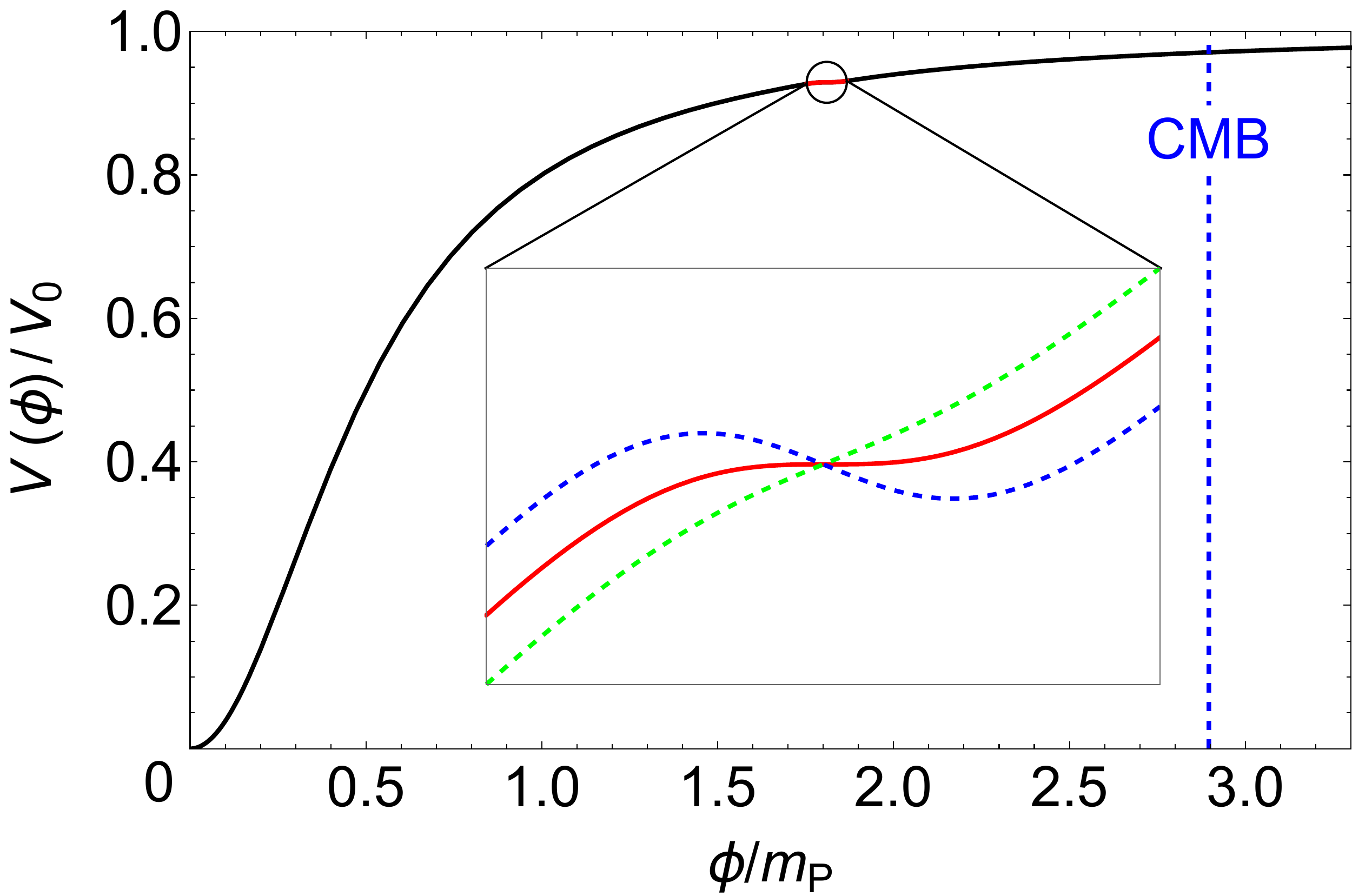}
\caption{The background inflaton potential $V_{\rm b}(\phi)$ in Eq. (\ref{kklt}) and the plateau (i.e., the USR region) caused by the perturbation $\dt V(\phi)$ in Eq. (\ref{3para}), with $\phi_0/\m=1.81$ and $\sigma/\m=0.057$. For comparison, $A$ is chosen to be $0.5V_{{\rm b},\phi}(\phi_0)$ (green dashed line), $V_{{\rm b},\phi}(\phi_0)$ (red line), and $1.5V_{{\rm b},\phi}(\phi_0)$ (blue dashed line), respectively. The CMB pivot scale $k_\ast$ corresponds to $\phi_\ast/\m=2.90$.} \label{fig:1-1}
\end{figure}

Below, the background inflaton potential $V_{\rm b}(\phi)$ is taken as the Kachru--Kallosh--Linde--Trivedi potential \cite{KKLT},
\begin{align}
V_{\rm b}(\phi)=V_0\f{\phi^2}{\phi^2+(m_{\rm P}/2)^2}, \label{kklt}
\end{align}
where we set $V_0/m_{\rm P}^4=10^{-10}$, so there can be a nearly scale-invariant power spectrum $\cP_\cR(k)$ on large scales and a relatively small tensor-to-scalar ratio $r$ favored by the CMB observations \cite{Planck}. Furthermore, $F$ is chosen to be Lorentzian function, so the perturbation $\dt V(\phi)$ is
\begin{align}
\dt V(\phi)=-A\f{(\phi-\phi_0)}{1+(\phi-\phi_0)^2/(2\sigma^2)}. \label{3para}
\end{align}
Altogether, the inflaton potential reads $V(\phi)=V_{\rm b}(\phi)+\dt V(\phi)$, as plotted in Fig. \ref{fig:1-1}.

\begin{figure}[htb]
\centering \includegraphics[width=0.65\linewidth]{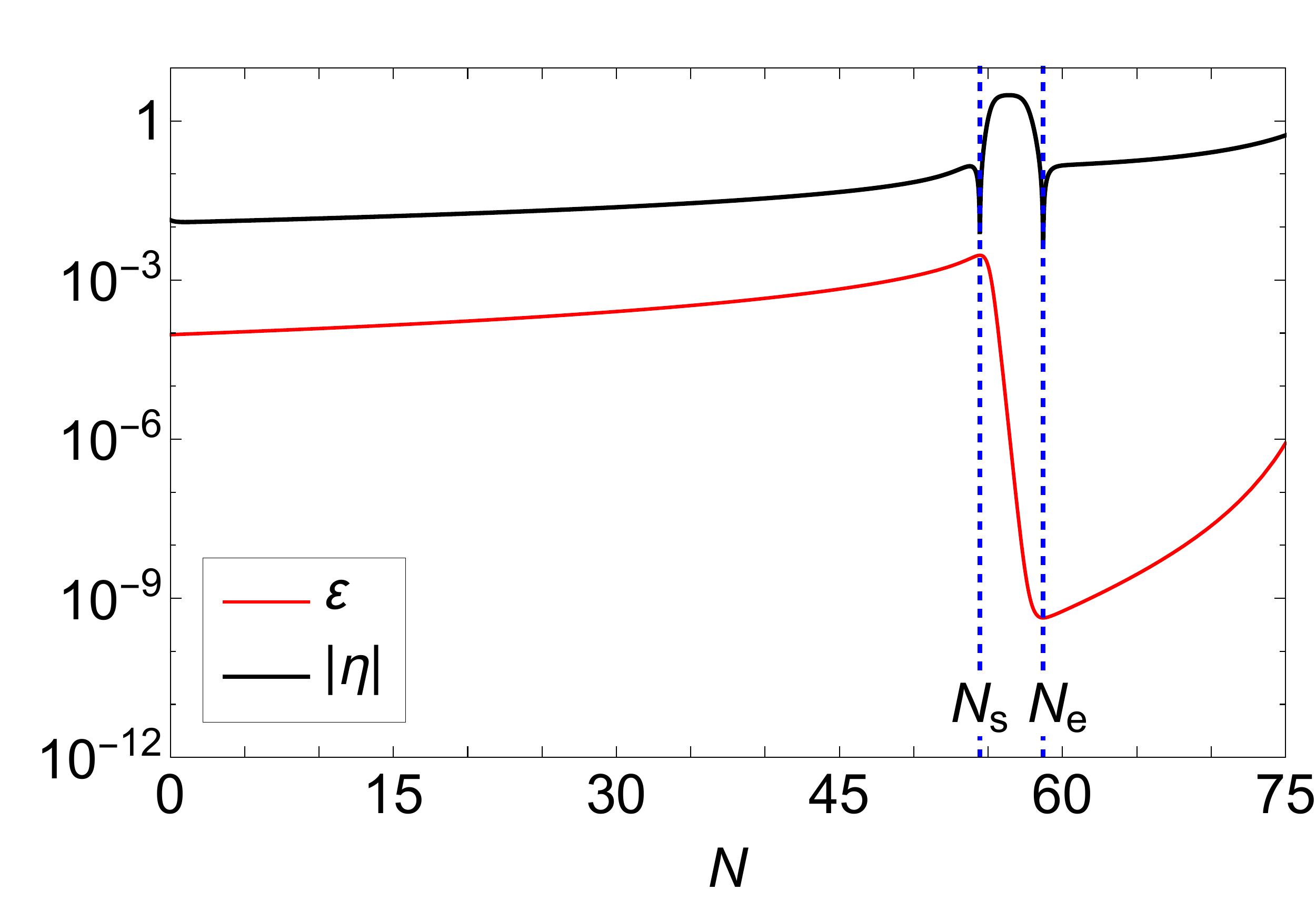}
\caption{The parameters $\ve$ and $\eta$ in the SR and USR stages, with the same model parameters as those in Fig. \ref{fig:1-1}. The starting and ending points of the USR stage are defined by $\eta(N_{\rm s})=\eta(N_{\rm e})=0$, with $N_{\rm s}=54.45$ and $N_{\rm e}=58.70$. During the USR stage, $\eta$ is negative and maintains around ${\cal O}(1)$; $\ve$ decreases exponentially and leads to a remarkable increase of the power spectrum $\cP_\cR(k)$, to be shown in Fig. \ref{fig:1PR}.} \label{fig:2-1}
\end{figure}

The plateau in $V_{\rm b}(\phi)$ caused by $\dt V(\phi)$ leads inflation to the USR stage, during which the inflaton field alters extremely slowly, so the parameters $\ve$ and $\eta$ change significantly. First, $\eta$ becomes positive, with the value maintaining around ${\cal O}(1)$. Hence, we can define the starting and ending points of the USR stage by $\eta(N_{\rm s})=\eta(N_{\rm e})=0$, with $N_{\rm s}$ and $N_{\rm e}$ being the corresponding numbers of $e$-folds. Second, $\ve$ decreases exponentially and induces a remarkable increase of the power spectrum $\cP_\cR(k)$. Altogether, the SR conditions are both broken in the USR stage, and the evolutions of $\ve$ and $\eta$ are shown in Fig. \ref{fig:2-1}.

\begin{figure}[htb]
\centering \includegraphics[width=0.65\linewidth]{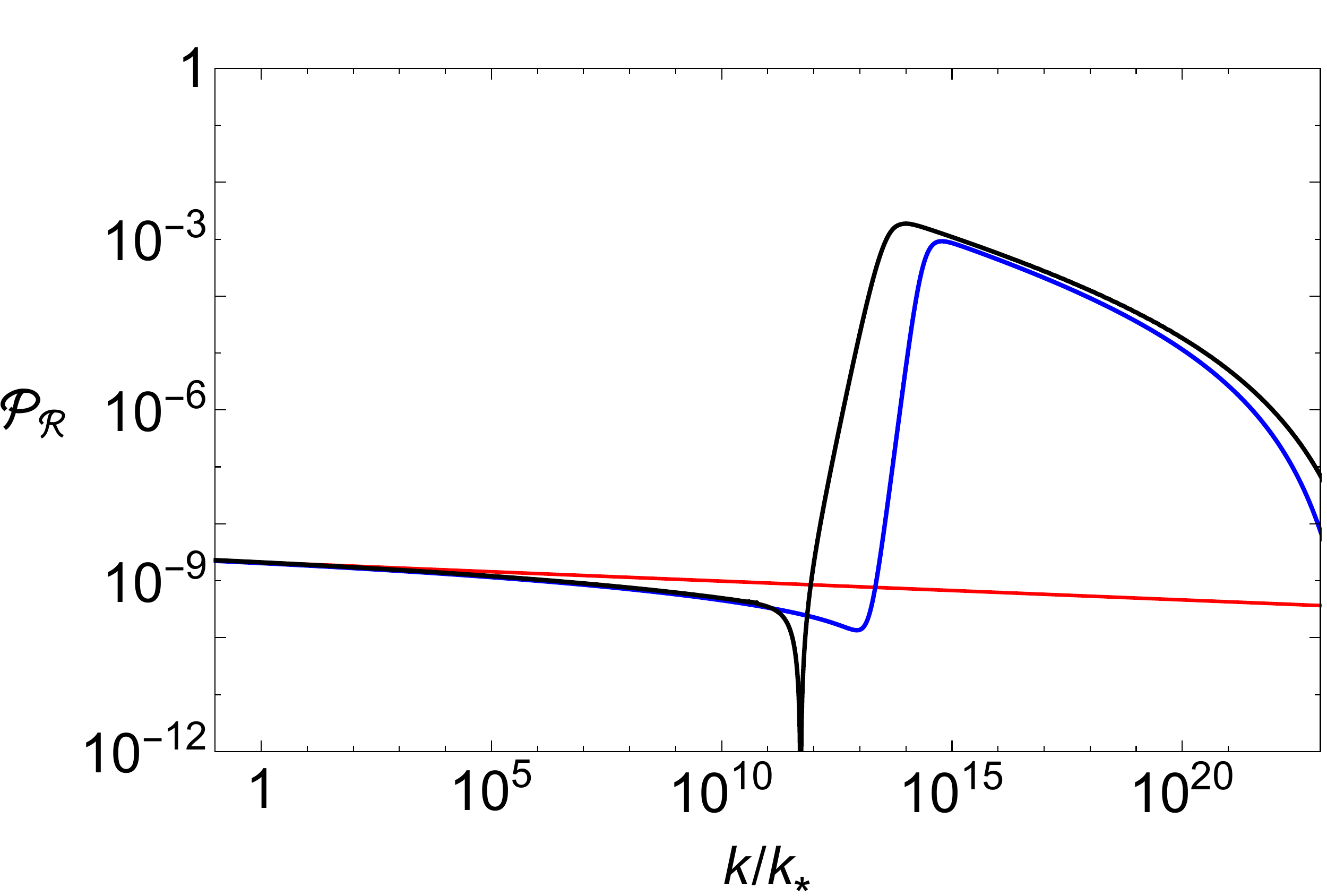}
\caption{The power spectra ${\cal P}_{\cal R}(k)$ influenced by the perturbation $\dt V(\phi)$ in Eq. (\ref{3para}): the power spectrum by numerically solving Eq. (\ref{MS}) (black line), the power spectrum from the SR approximations in Eq. (\ref{PRsr}) (blue line), and the nearly scale-invariant power spectrum for comparison (red line), with the same model parameters as those in Fig. \ref{fig:1-1}. The SR power spectrum significantly deviates from the numerical result, for both the position and height of its peak, and can no longer be trusted in the USR inflation.} \label{fig:1PR}
\end{figure}

Furthermore, the influences of $\dt V(\phi)$ on the power spectra ${\cal P}_{\cal R}(k)$ are shown in Fig. \ref{fig:1PR}, where we plot the power spectra both by numerically solving Eq. (\ref{MS}) and from the SR approximations in Eq. (\ref{PRsr}). In addition, the nearly scale-invariant power spectrum is also shown for comparison.\footnote{Here, we should emphasize that it is the scales that exit the horizon before the USR stage that account for the sharp rise in ${\cal P}_{\cal R}(k)$. This is because, in the SR stage, ${\cal R}_{k,N}$ decreases exponentially outside the horizon, so ${\cal R}_k$ is almost constant; in contrast, in the USR stage, ${\cal R}_{k,N}$ still increases outside the horizon, so ${\cal R}_k$ can continue increasing. Consequently, there is a competition between the decrease and increase of ${\cal R}_{k,N}$, and they balance at a characteristic scale $\tilde{k}$ \cite{liu}. Therefore, for the scales with $\tilde{k}<k<k(N_{\rm s})$ (i.e., the scales exiting the horizon before the USR stage but still in the SR stage), the increase of ${\cal R}_{k,N}$ in the USR stage can exceed the decrease of ${\cal R}_{k,N}$ in the SR stage, leading to the sharp rise in ${\cal P}_{\cal R}(k)$. This also explains the offset of the steep growth of ${\cal P}_{\cal R}(k)$ between the numerical result and the SR approximations [i.e., the numerical ${\cal P}_{\cal R}(k)$ rises up at smaller $k$].} The SR power spectrum significantly deviates from the numerical result, for both the position and height of its peak. Meanwhile, the PBH abundance is highly sensitive to these two factors, so the SR approximations cannot be trusted any longer in the USR stage, and all relevant calculations below are performed numerically.

Now, we calculate the power spectra ${\cal P}_{\cal R}(k)$ and demand the PBH abundances $f(M)\sim 0.1$, with the PBH masses $M$ in the three typical mass windows at $10^{-17}M_\odot$, $10^{-13}M_\odot$, and $30M_\odot$, respectively. In this process, we have three model parameters $A$, $\phi_0$, and $\sg$ but only two constraints from $M$ and $f$, so we will be confronted with the parameter degeneracy. Therefore, in this section, we further impose a restriction as $A=V_{{\rm b},\phi}(\phi_0)$ for convenience, meaning that the perturbation $\dt V(\phi)$ can create a perfect plateau at $\phi_0$. Thus, our task is reduced to determining the remaining two parameters $\phi_0$ and $\sg$. Moreover, the parameter degeneracy can simultaneously result in slightly different scalar spectral index $n_{\rm s}$ and tensor-to-scalar $r$, and the bounds $n_{\rm s}=0.965\pm0.008$ and $r<0.06$ ($95\%$ confidence level) \cite{Planck} must be taken into account in parameter adjustment.

\begin{figure*}[htb]
\begin{center}
\includegraphics[width=0.45\linewidth]{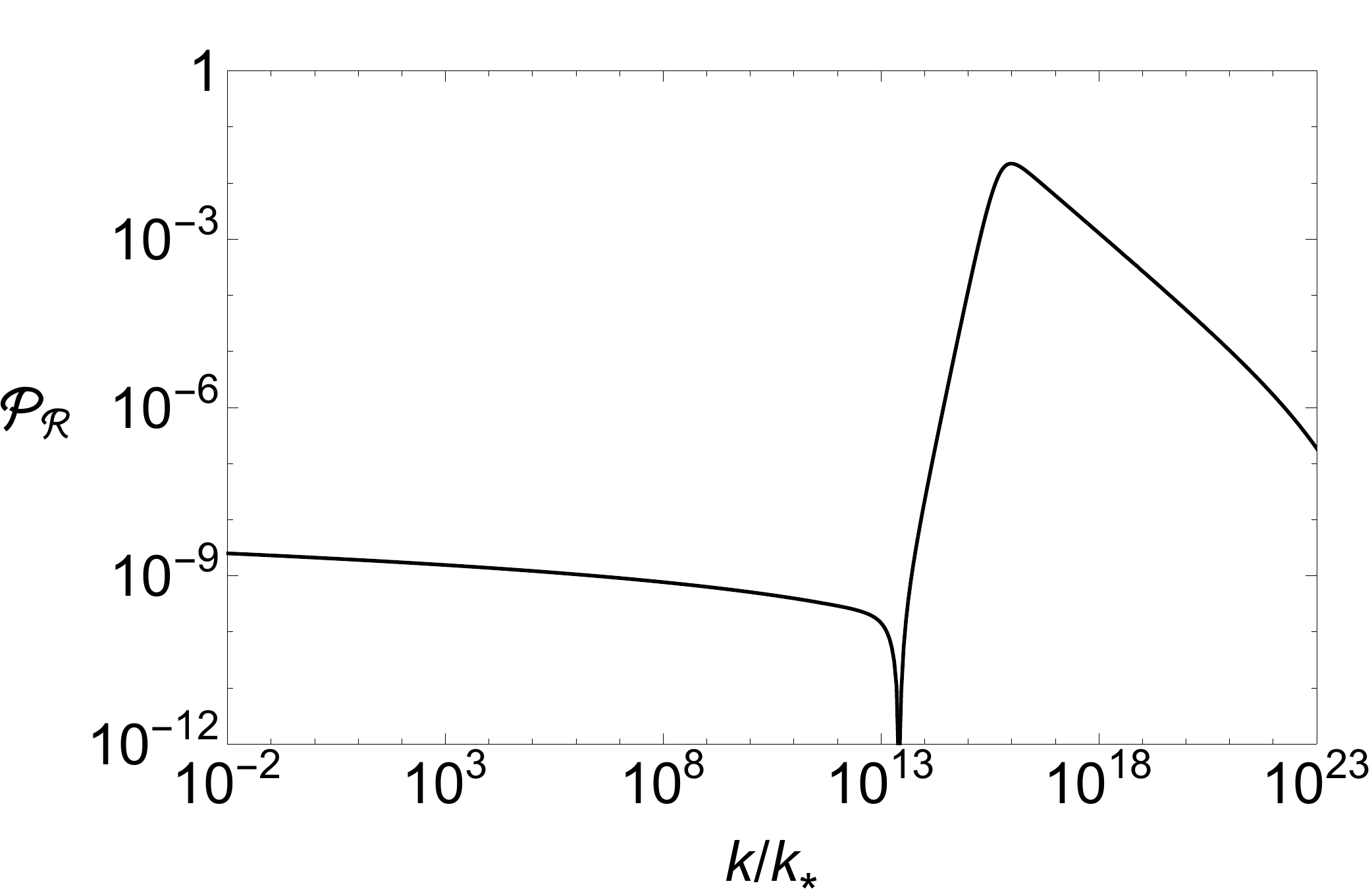} \quad \includegraphics[width=0.45\linewidth]{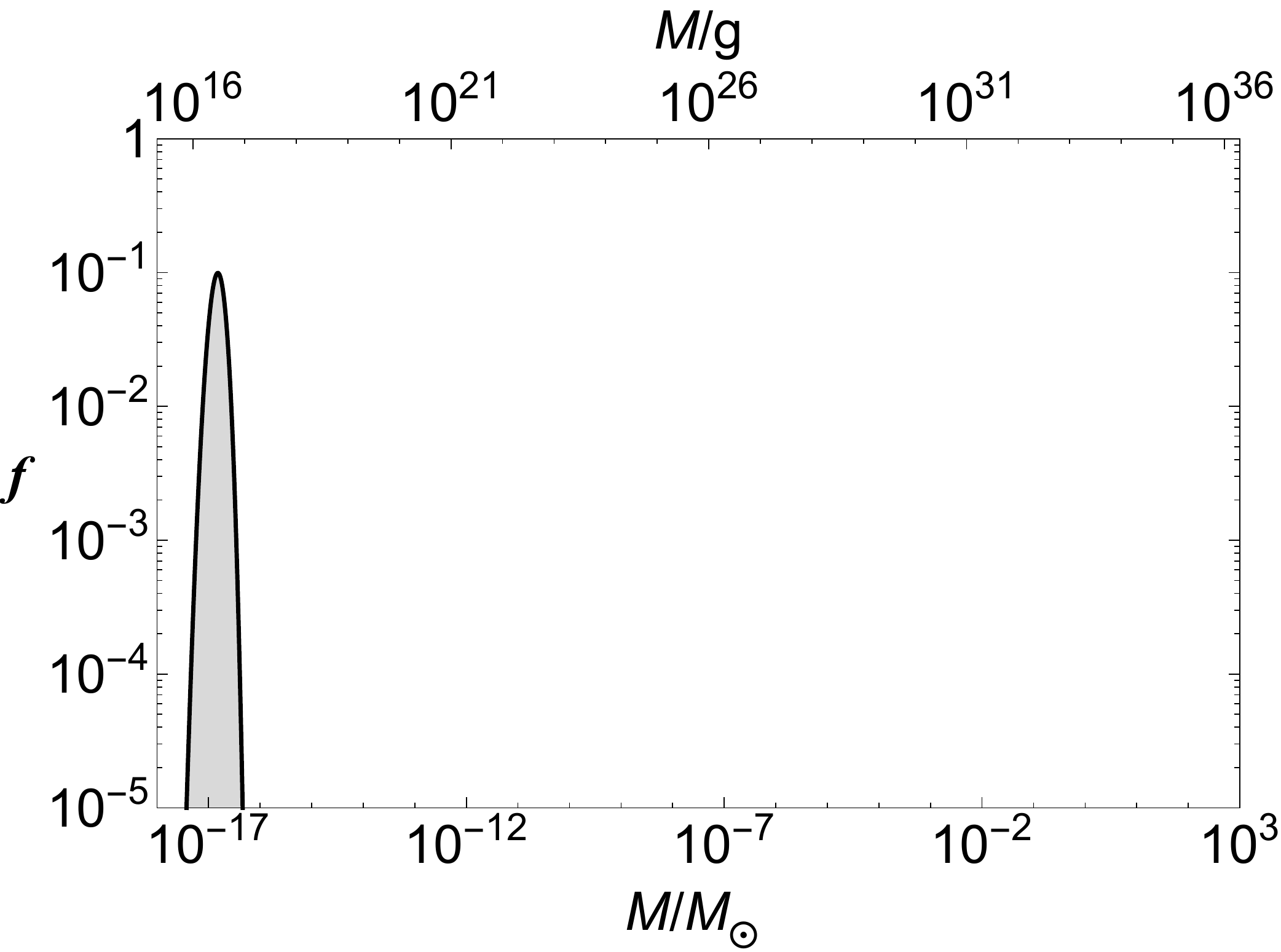} \\ \vskip .1cm
\includegraphics[width=0.45\linewidth]{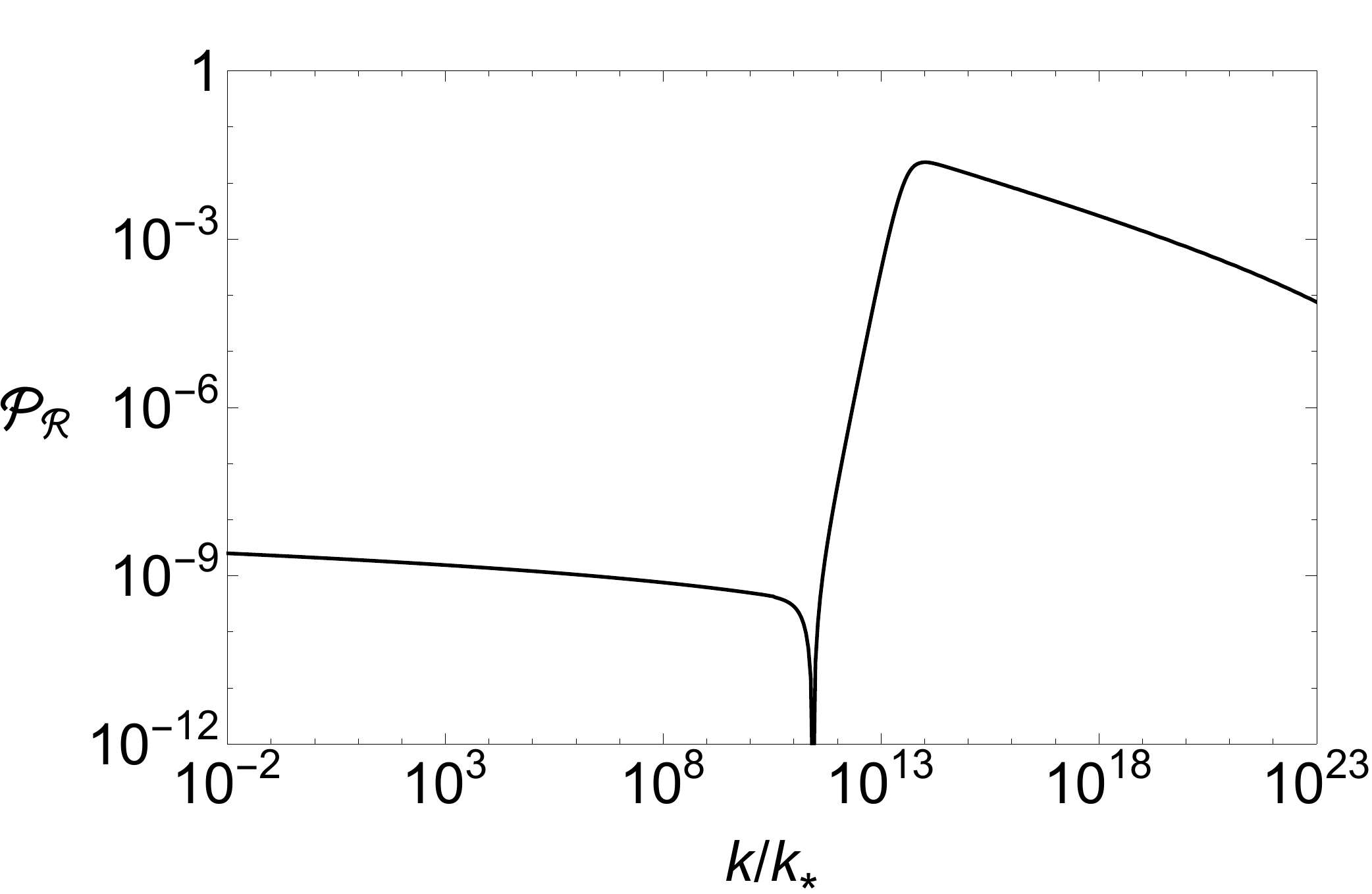} \quad \includegraphics[width=0.45\linewidth]{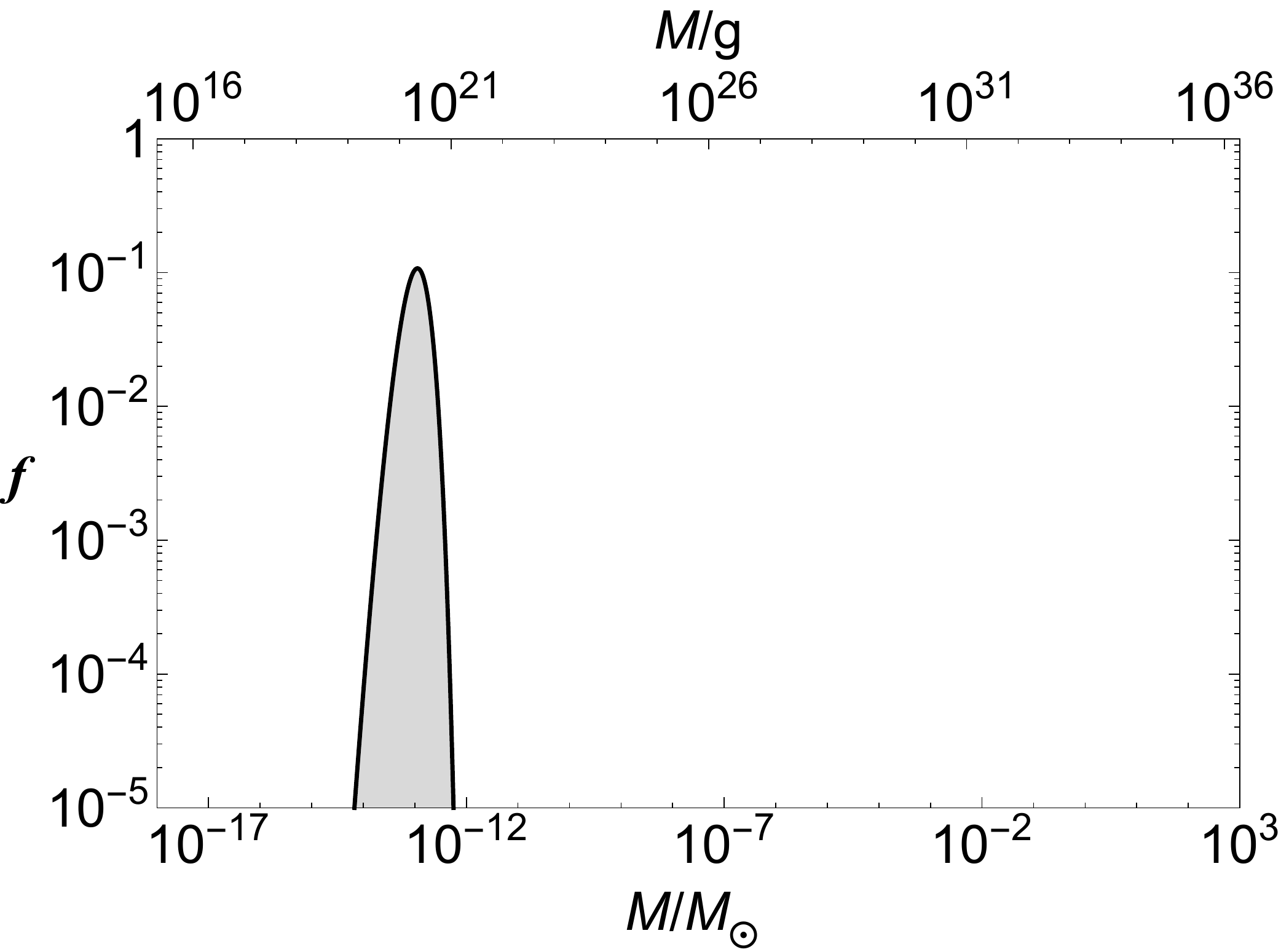} \\ \vskip .1cm
\includegraphics[width=0.45\linewidth]{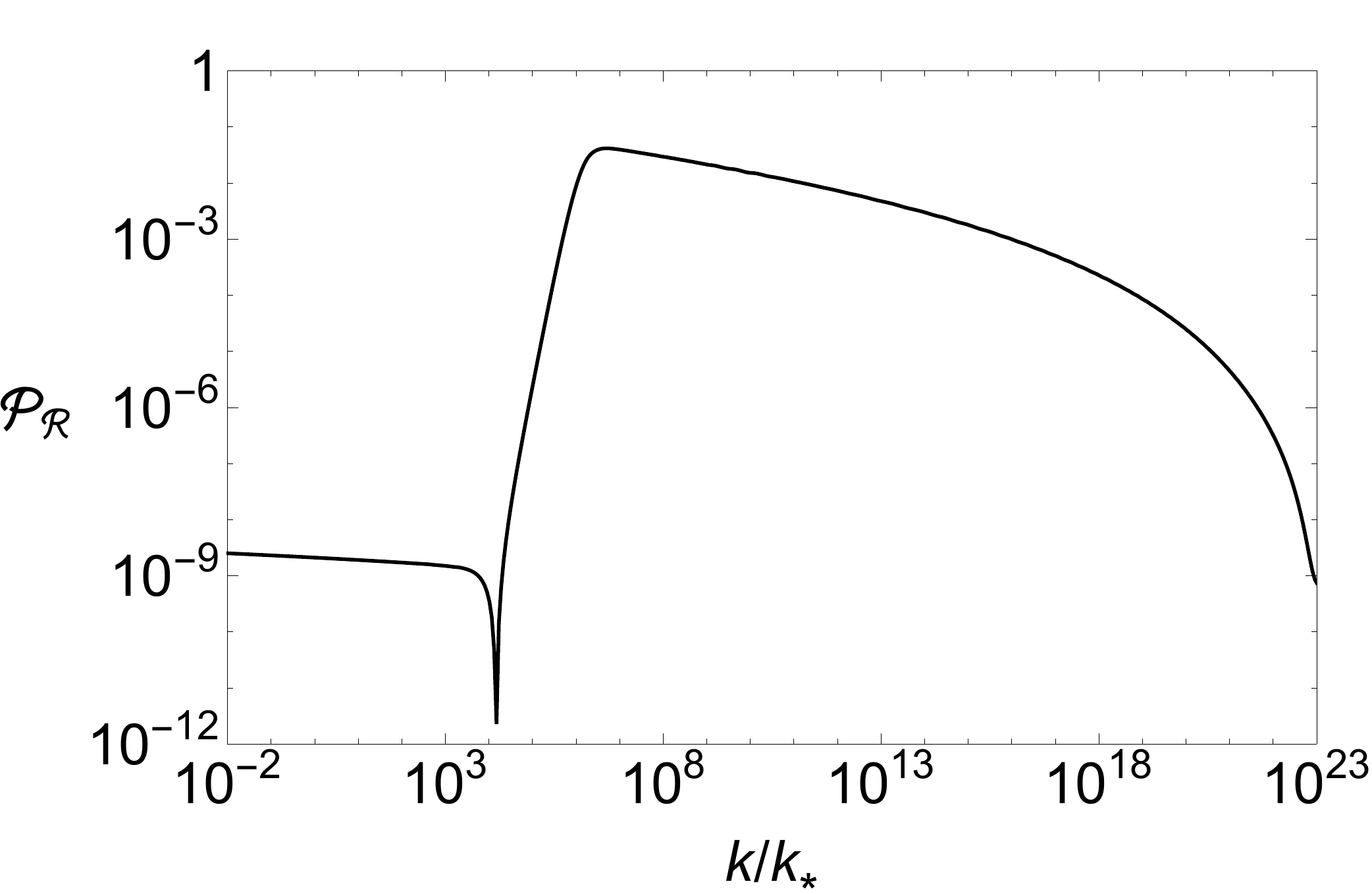} \quad \includegraphics[width=0.45\linewidth]{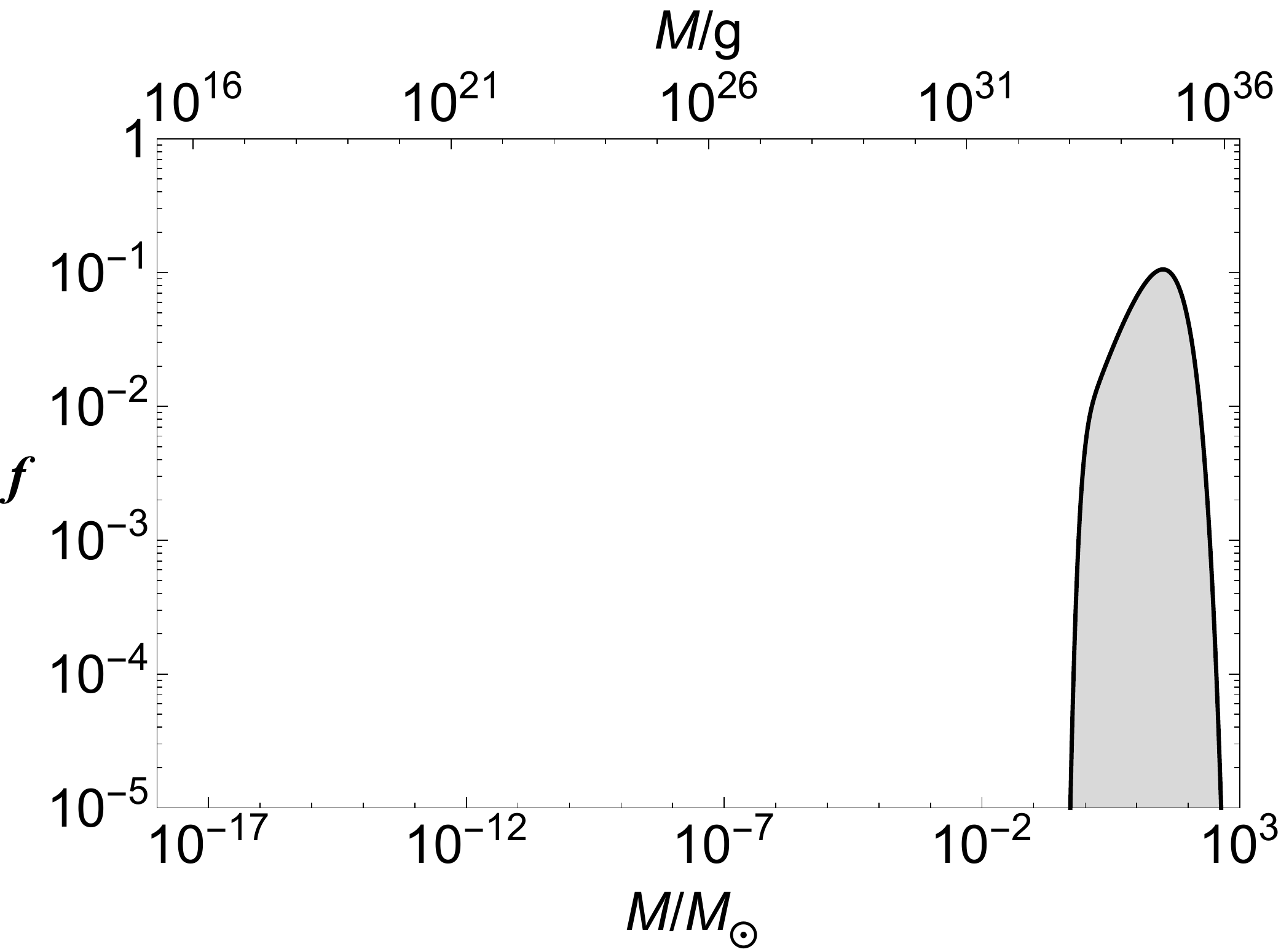}
\end{center}
%\begin{center}
%\includegraphics[width=0.3\linewidth]{17-1-1.pdf} \quad
%\includegraphics[width=0.3\linewidth]{13-1-1.pdf} \quad
%\includegraphics[width=0.3\linewidth]{30-1-1.pdf} \\ \vskip .1cm
%\includegraphics[width=0.3\linewidth]{17-2-1.pdf} \quad
%\includegraphics[width=0.3\linewidth]{13-2-1.pdf} \quad
%\includegraphics[width=0.3\linewidth]{30-2-1.pdf}
%\end{center}
\caption{The power spectra ${\cal P}_{\cal R}(k)$ and the relevant PBH abundances $f(M)$ with the PBH masses $M$ in the three typical mass windows at $10^{-17}M_\odot$, $10^{-13}M_\odot$, and $30M_\odot$, respectively. If $f\sim 0.1$, ${\cal P}_{\cal R}$ needs to be enhanced up to at least $10^{-2}$ on small scales, seven orders of magnitude higher than its value on large scales. From Eq. (\ref{M}), with $M$ increasing, $k_\pb$ decreases, so the peak of ${\cal P}_{\cal R}(k)$ moves to larger scales.} \label{fig:one6}
\end{figure*}

The power spectra ${\cal P}_{\cal R}(k)$ and the relevant PBH abundances $f(M)$ are plotted in Fig. \ref{fig:one6}, and the corresponding parameters $\phi_0$ and $\sg$ are summarized in Tab. \ref{tab:1}. We set the initial conditions for inflation to be $\phi/\m=3.30$ and $\phi_{,N}/\m=-0.0137$, and we have $n_{\rm s}=0.9591$ and $r=0.00321$ ($10^{-17}M_\odot$), $n_{\rm s}=0.9588$ and $r=0.00322$ ($10^{-13}M_\odot$), and $n_{\rm s}=0.9580$ and $r=0.00322$ ($30M_\odot$), all in their $95\%$ confidence levels \cite{Planck}.

\begin{table}[htb]
\renewcommand\arraystretch{1.25}
\centering
\begin{tabular}{m{1.2cm}<{\centering}|m{1.2cm}<{\centering}|m{1.6cm}<{\centering}}
\hline\hline
$M/M_\odot$ & $\phi_0/m_{\rm P}$ & $\sigma/m_{\rm P}$ \\
  \hline
$10^{-17}$ &  1.31  &  0.0831881  \\
  \hline
$10^{-13}$ &  1.81  &  0.0405471  \\
\hline
$30$       &  2.56  &  0.0159387  \\
  \hline\hline
\end{tabular}
\caption{The parameters $\phi_0$ and $\sg$ for the PBH abundances $f\sim 0.1$ in the three typical mass windows at $10^{-17}M_\odot$, $10^{-13}M_\odot$, and $30M_\odot$, respectively. With $M$ increasing, $\phi_0$ increases (the USR stage occurs earlier) and $\sg$ decreases (the duration of the USR stage shortens). Moreover, $f$ is highly sensitive to $\sigma$.} \label{tab:1}
\end{table}

From Fig. \ref{fig:one6} and Tab. \ref{tab:1}, our basic results can be summarized as follows. First, if the PBH abundances $f$ are around $0.1$, the power spectra ${\cal P}_{\cal R}$ should be enhanced up to at least $10^{-2}$ on small scales, seven orders of magnitude higher than the value on large scales (e.g., ${\cal P}_{\cal R}\sim 10^{-9}$ on the CMB pivot scale $k_\ast$). Second, with the PBH mass $M$ increasing, the peak of ${\cal P}_{\cal R}$ moves to larger scales, as can be seen from Eq. (\ref{M}) that a larger $M$ corresponds to a smaller $k_\pb$. Moreover, a smaller $k_\pb$ means an earlier USR stage, so the parameter $\phi_0$ increases with $M$, as shown in Tab. \ref{tab:1}. Third, a larger $\phi_0$ would enhance ${\cal P}_{\cal R}$, so if the height of ${\cal P}_{\cal R}$ maintains around $10^{-2}$, the parameter $\sg$ must decrease accordingly. This is because the earlier the USR inflation occurs (with larger $\phi_0$), the more slowly the inflaton rolls down. Therefore, the width $\sg$ of the plateau must become smaller (with shorter duration of the USR stage), so that the influence of $\dt V(\phi)$ would not be too large, as also shown in Tab. \ref{tab:1}. Last, the precision of $\sg$ is notably high, with at least six significant digits. This is expected, as from Eqs. (\ref{zheteng}) and (\ref{GLMS}), the PBH abundance $f(M)$ is exponentially sensitive to $\sg$, so the step size of $\sg$ must be as small as $10^{-7}\m$ in parameter adjustment.

\section{PBHs from two perturbations in the inflaton potential} \label{sec:two}

In this section, we further investigate the cases with two perturbations in the inflaton potential in the GLMS approximation, in which there can be PBHs of different masses both with the abundance $f\sim 0.1$ in two of the three typical mass windows at $10^{-17}M_\odot$, $10^{-13}M_\odot$, and $30M_\odot$.

Below, the $F$ function in Eq. (\ref{F}) is chosen to be Gaussian,
\begin{align}
\dt V(\phi)=-A(\phi-\phi_0)\exp\lt[-\frac{(\phi-\phi_0)^2}{2\sg^2}\rt], \label{Gauss}
\end{align}
and the reasons are twofold. One is to verify the applicability of our model by simply taking the perturbation in another form; the other and more essential one is to avoid the interference between the two perturbations. First, the two perturbations cannot be too far away; otherwise, the inflaton will spend more time on the first plateau on large scale and will pass the second plateau on small scale much later, making the relevant PBH mass extremely small. Second, the two perturbations cannot be too close either; otherwise, there will be strong parameter degeneracy. Consequently, the two perturbations should be separated at a moderate distance, and at the same time the two plateaus should be narrower than before. Compared to the Lorentzian form in Eq. (\ref{3para}), the Gaussian form in Eq. (\ref{Gauss}) converges more quickly and can avoid the interference more effectively. Now, the inflaton potential reads $V(\phi)=V_{\rm b}(\phi)+\dt V_1(\phi)+\dt V_2(\phi)$.

In the current case with two perturbations, it is impossible to set $A=V_{{\rm b},\phi}(\phi_0)$ for each perturbation any longer. Therefore, we re-parameterize $A$ in the form of $A=V_{{\rm b},\phi}(\phi_0)(1+A_0)$, with $A_0$ characterizing the deviation of the inflaton potential from a perfect plateau at $\phi_0$. Altogether, there are six model parameters: $A_0^{(1)}$, $\phi_0^{(1)}$, $\sigma^{(1)}$, $A_0^{(2)}$, $\phi_0^{(2)}$, and $\sigma^{(2)}$ (the superscripts 1 and 2 stand for small and large PBH masses, respectively).

\begin{figure*}[htb]
\centering
\includegraphics[width=0.45\linewidth]{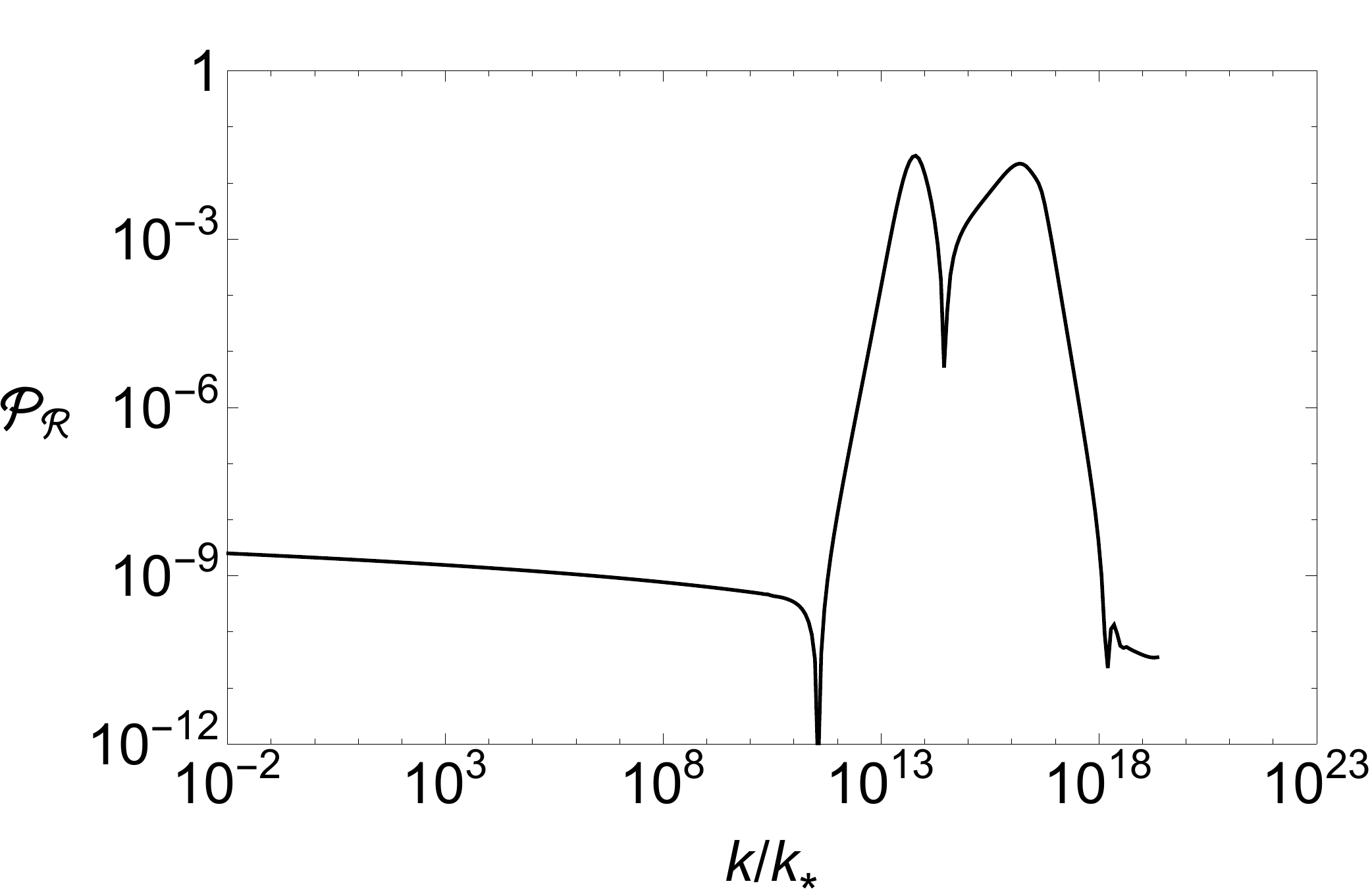} \quad \includegraphics[width=0.45\linewidth]{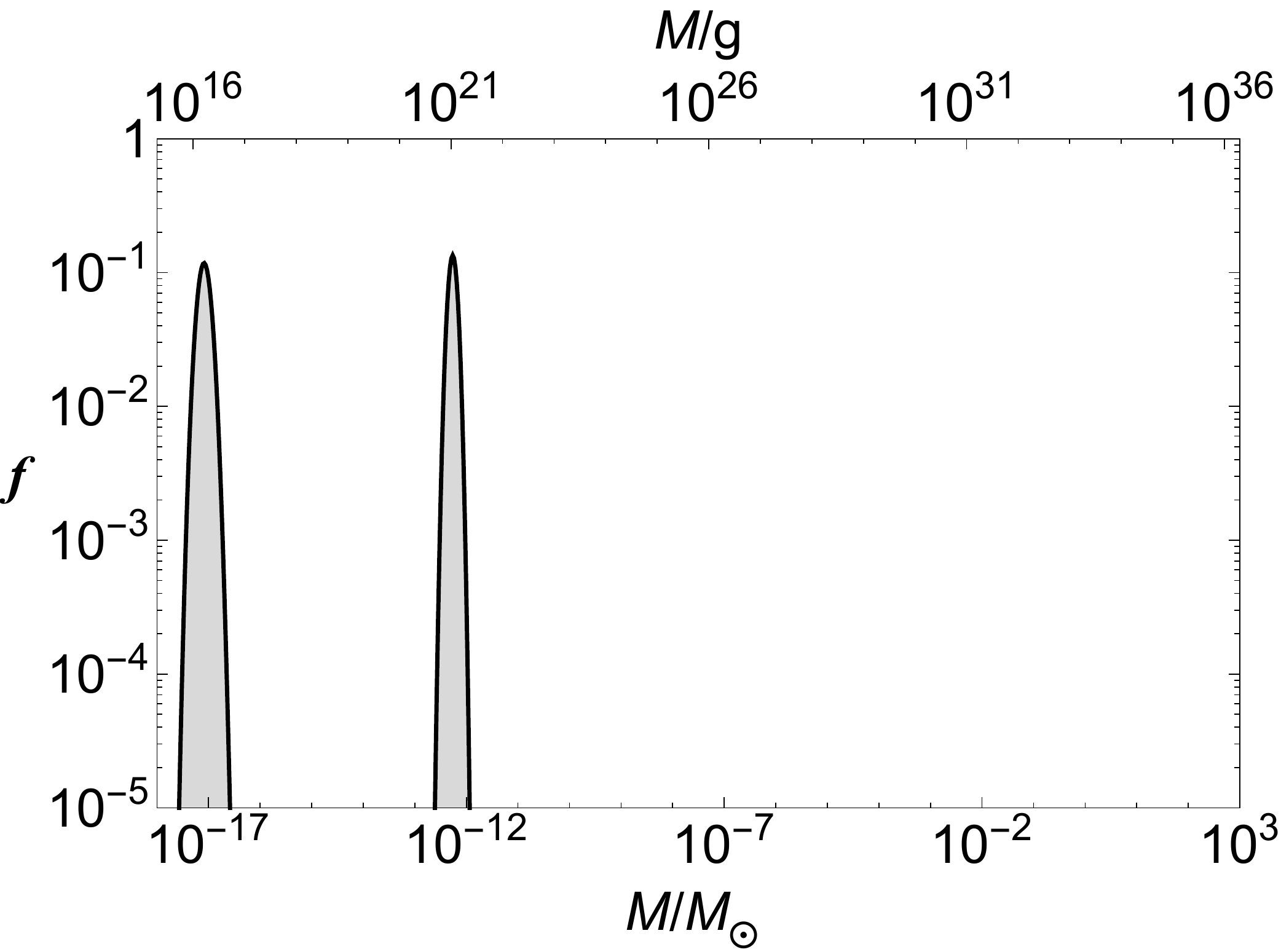} \\ \vskip .1cm
\includegraphics[width=0.45\linewidth]{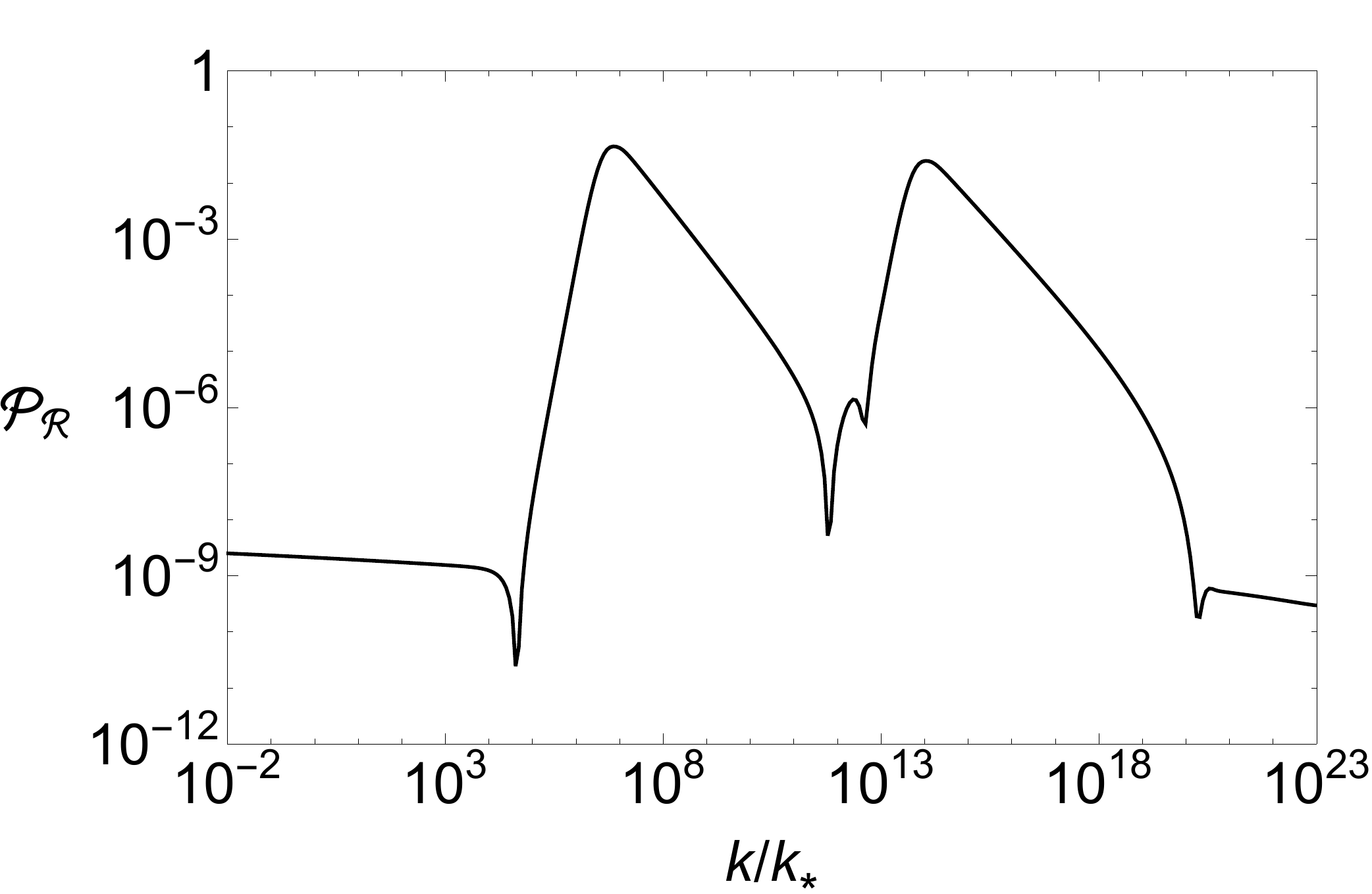} \quad \includegraphics[width=0.45\linewidth]{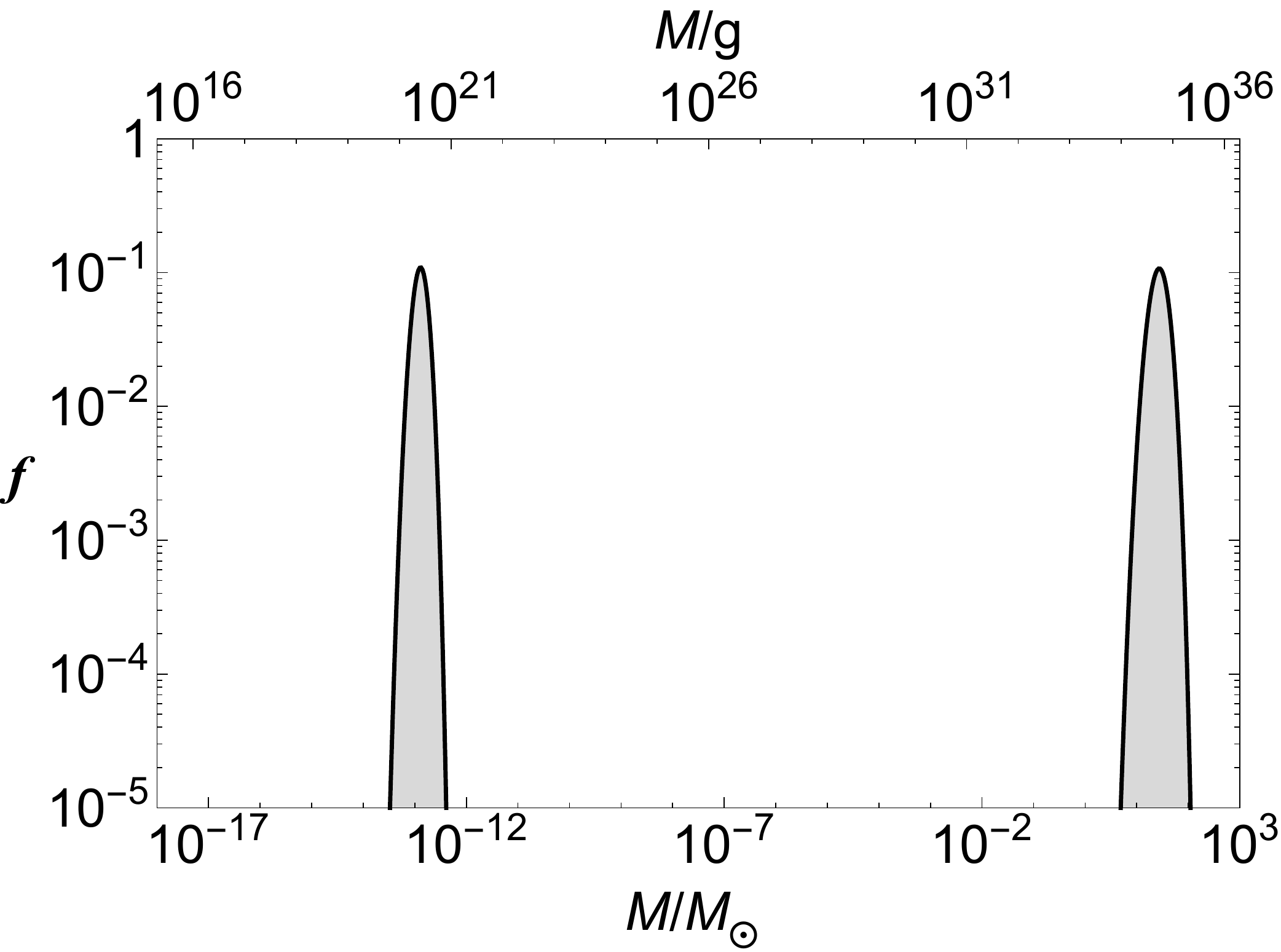} \\ \vskip .1cm
\includegraphics[width=0.45\linewidth]{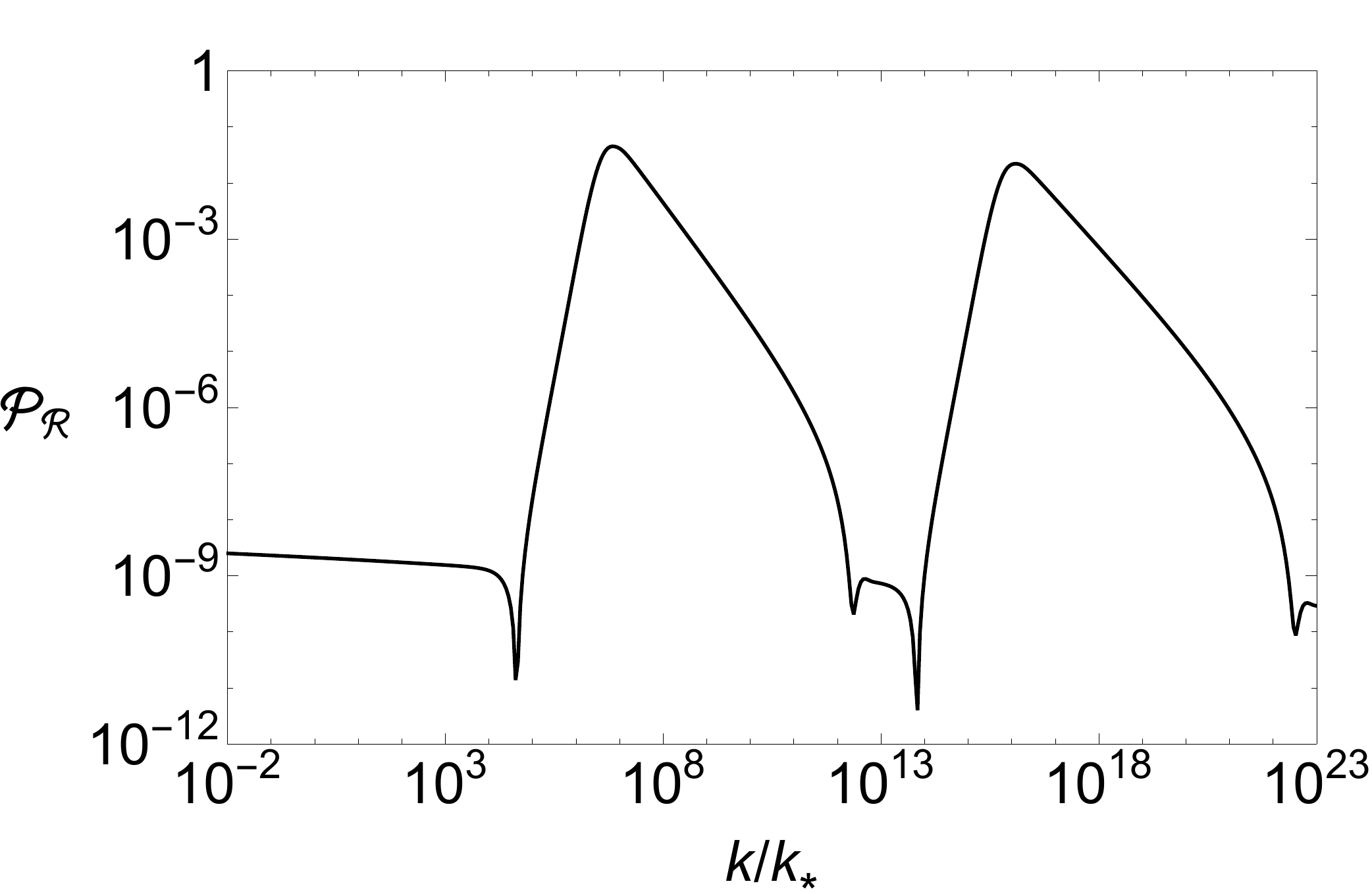} \quad \includegraphics[width=0.45\linewidth]{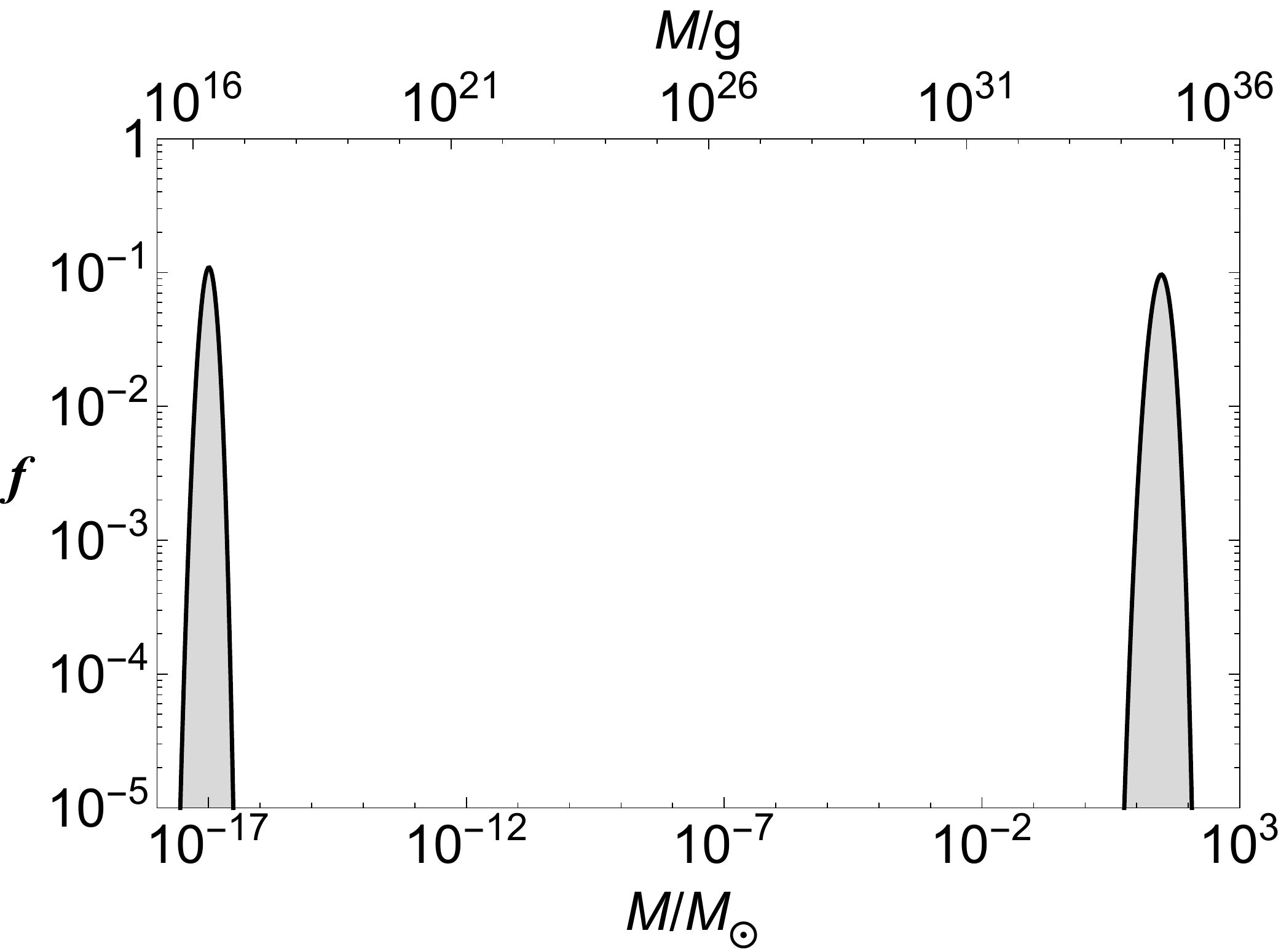}
\caption{The power spectra ${\cal P}_{\cal R}(k)$ and the relevant PBH abundances $f(M)$ with the PBHs of different masses in two of the three typical mass windows at $10^{-17}M_\odot$, $10^{-13}M_\odot$, and $30M_\odot$. The interference between the two perturbations in the inflaton potential induces a turning point in the growing stage of ${\cal P}_{\cal R}(k)$ on the small scale (middle left panel). This turning point tends to disappear when the two perturbations are far away (lower left panel) and also disappears when the two perturbations are too close, but with a strong distortion in ${\cal P}_{\cal R}(k)$ at the same time (upper left panel).} \label{fig:two6}
\end{figure*}

According to the separation between the two PBH masses, the power spectra ${\cal P}_{\cal R}(k)$ and the relevant PBH abundances $f(M)$ are plotted in Fig. \ref{fig:two6}, and the corresponding model parameters are summarized in Tab. \ref{tab:2}. The initial conditions for inflation are kept the same as those in Sec. \ref{sec:one}, and the changes of the resulting $n_{\rm s}$ and $r$ are negligible.

\begin{table*}[htb]
\renewcommand\arraystretch{1.25}
\centering
\resizebox{1\columnwidth}{!}{
\begin{tabular}{m{2.8cm}<{\centering}| m{1.5cm}<{\centering}| m{1.3cm}<{\centering}| m{1.5cm}<{\centering}| m{1.5cm}<{\centering}| m{1.3cm}<{\centering}| m{1.5cm}<{\centering}| m{2.7cm}<{\centering}}
\hline\hline
$M/M_\odot$ & $A_0^{(1)}$ & $\phi_0^{(1)}/\m$ & $\sigma^{(1)}/\m$ & $A_0^{(2)}$ & $\phi_0^{(2)}/\m$ & $\sigma^{(2)}/\m$ & $(\phi_0^{(2)}-\phi_0^{(1)})/\m$ \\
  \hline
$10^{-17}$ and $10^{-13}$ & 0.2736210 & 1.700 & 0.029705  & 0.1383814 & 1.810 & 0.029002  & 0.110\\
  \hline
$10^{-13}$ and $30$       & 0.0121025 & 2.377 & 0.0200459 & 0.0160910 & 2.521 & 0.0164554 & 0.144\\
  \hline
$10^{-17}$ and 30         & 0.0119941 & 2.165 & 0.0260763 & 0.0180321 & 2.522 & 0.0162205 & 0.357\\
\hline\hline
\end{tabular}}
\caption{The parameters $A_0^{(1)}$, $\phi_0^{(1)}$, $\sigma^{(1)}$, $A_0^{(2)}$, $\phi_0^{(2)}$, and $\sigma^{(2)}$ (the superscripts 1 and 2 stand for small and large PBH masses, respectively) for the PBHs of different masses both with the abundance $f\sim 0.1$ in two of the three typical mass windows at $10^{-17}M_\odot$, $10^{-13}M_\odot$, and $30M_\odot$. With the separation between the two mass windows decreasing, the separation between the two perturbations $\Delta\phi= \phi_0^{(2)}-\phi_0^{(1)}$ also decreases, but $A_0$ increases, so as to avoid the interference between the two USR stages.} \label{tab:2}
\end{table*}

From Fig. \ref{fig:two6} and Tab. \ref{tab:2}, we arrive at the following results. First, analogous to the case with one perturbation, if the PBH abundances $f$ are around $0.1$ in two mass windows simultaneously, the power spectra $\cP_\cR(k)$ must also possess two peaks both with the height of $10^{-2}$. Second, from the first and last columns in Tab. \ref{tab:2}, we observe that the separation between the two perturbations $\Delta\phi=\phi_0^{(2)}-\phi_0^{(1)}$ increases with that between the two mass windows, and this tendency also applies to the separation between the two peaks in $\cP_\cR(k)$. Third, as $\Delta\phi$ decreases, the parameter $A_0$ increases. This is because, when the two perturbations are closer, we must be very cautious to avoid their interference. However, this cannot be done simply by decreasing the width $\sg$ of the perturbation, because if $\sg$ is too small, the USR stage cannot last long enough to enhance $\cP_\cR(k)$ to $10^{-2}$ as needed. Therefore, a larger $A_0$ is indispensable, so as to incline the plateau and to play a similar role of a smaller $\sg$. Fourth, we point out an interesting feature in ${\cal P}_{\cal R}(k)$ in Fig. \ref{fig:two6}: the turning point in the growing stage of ${\cal P}_{\cal R}(k)$ on the small scale. This is a natural consequence of the interference between the two perturbations, which induces an overlap between the decaying stage of ${\cal P}_{\cal R}(k)$ on the large scale and the growing stage of ${\cal P}_{\cal R}(k)$ on the small scale. This turning point can be clearly seen, if the two perturbations are separated at a moderate distance (middle left panel). Moreover, if the two perturbations are far away, the turning point tends to disappear (lower left panel); if they are too close, the turning point also disappears, but there is a strong distortion in ${\cal P}_{\cal R}(k)$ (upper left panel).

% Last, we should mention that, in the inflation models with multiple USR phases, the steepest growth of ${\cal P}_{\cal R}(k)$ may be larger than the result $k^4$ in Refs. \cite{Byrnes:2018txb, liu}. It was found to be $k^5(\ln k)^2$ in Ref. \cite{Carrilho:2019oqg}, and the spectral index could be even enhanced to $n_{\rm s}-1=8$ for the second USR phase, indicating that $n_{\rm s}$ keeps memory of the history of the multiple USR inflation \cite{Tasinato:2020vdk}. Nevertheless, different from these analytical results, we perform all our calculations numerically in this paper. In Fig. \ref{fig:two6}, the discrepancy between the slopes of the two growing stages in ${\cal P}_{\cal R}(k)$ seems not to be so evident as that in Ref. \cite{Tasinato:2020vdk}. This is because we have adjusted the model parameters in Tab. \ref{tab:2} to avoid the interference between the two USR phases, so the memory effect of $n_{\rm s}$ is not so strong.

\section{Comparison of peak and PS theories} \label{sec:bj}

The PBH abundance $f$ depends not only on the power spectrum, but also on the specific methods to calculate the PBH mass fraction $\beta$. In Sec. \ref{sec:peak}, we have discussed three different approximations based on the general peak theory, according to the spectral moments involved. In Secs. \ref{sec:one} and \ref{sec:two}, we use only the GLMS approximation. Now, we compare all these three approximations in calculating the PBH abundance, with one perturbation in the inflaton potential.

\begin{figure}[htb]
\centering
\includegraphics[width=0.65\linewidth]{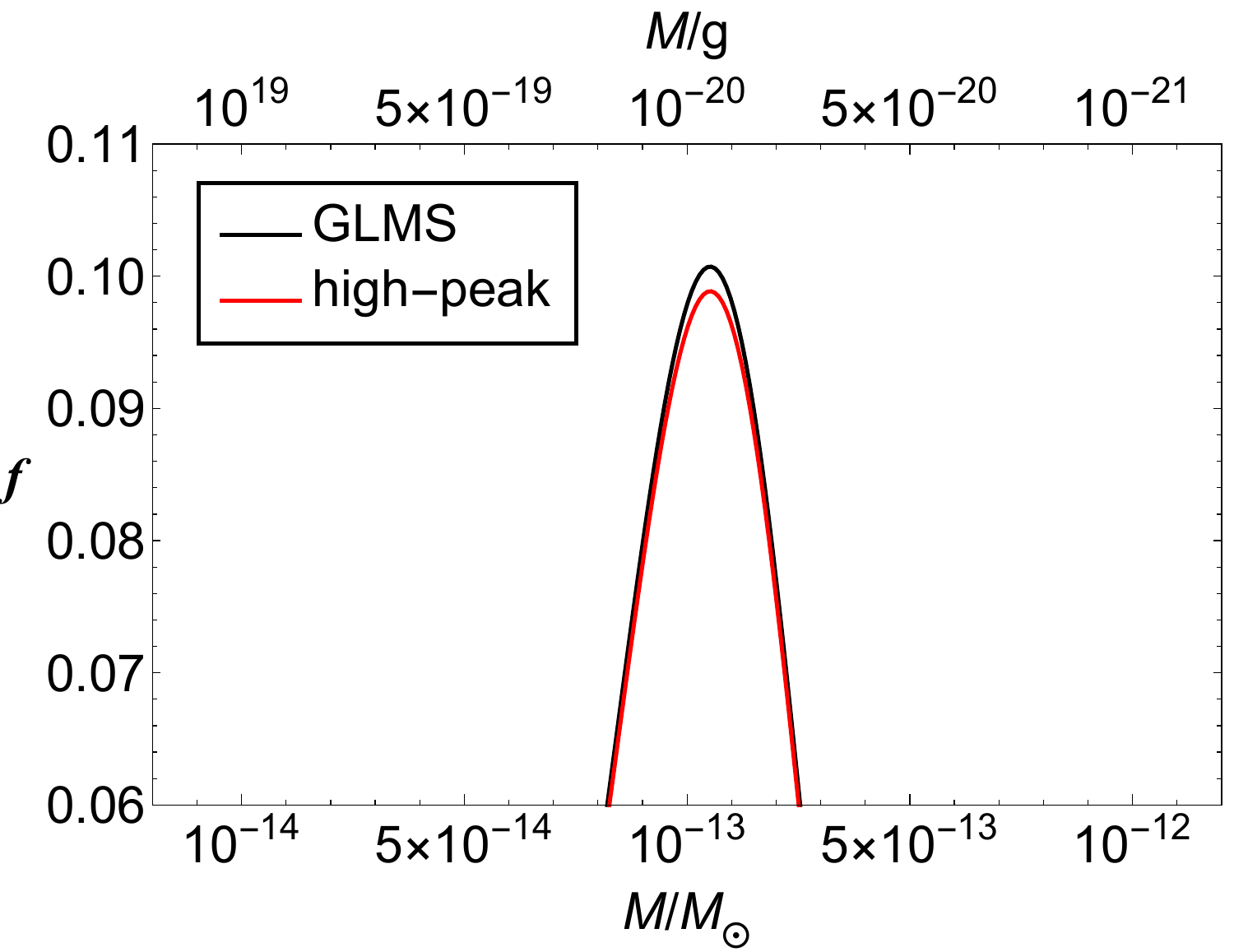}
\caption{The PBH abundances $f(M)$ obtained from the GLMS approximation (black line) and the high-peak approximation (red line), with the PBH mass being around $10^{-13}M_\odot$. The $f(M)$ from the GLMS approximation is set to be $0.1$, and the $f(M)$ from the high-peak approximation is $0.981428$. This slight difference is from the facts that $\nu_{\rm c}=8.82676 \gg 1$ and $\gamma=0.823356 \approx 1$, so the ratio of $\beta_{\rm hp}/\beta_{\rm GLMS}$ is pretty close to 1, consistent with Eqs. (\ref{hp}) and (\ref{GLMS}).} \label{fig:compare2}
\end{figure}

First, we compare the PBH abundances obtained from the high-peak and GLMS approximations, with the PBH mass being around $10^{-13}M_\odot$. As explained in Sec. \ref{sec:peak}, the PBH mass fraction $\beta_{\rm hp}$ is only slightly less than $\beta_{\rm GLMS}$ (too little to be distinguished, especially in the logarithmic coordinate system in Fig. \ref{fig:one6}), and the difference mainly stems from the $\gamma$ factor. In Fig. \ref{fig:compare2}, the PBH abundance obtained from the GLMS approximation is set to be $0.1$, with the relevant factors being $\nu_{\rm c}=8.82676$ and $\gamma=0.823356$. Therefore, from Eqs. (\ref{hp}) and (\ref{GLMS}),
\begin{align}
\f{\beta_{\rm hp}}{\beta_{\rm GLMS}}=\f{\nu_{\rm c}^2+2-3/\gamma^2}{\nu_{\rm c}^2-1}=0.981428. \n
\end{align}
The slight difference lies in the fact that $\nu_{\rm c}$ is much greater than 1, as the PBH formation is a rather rare event. This ratio indicates that the discrepancy between the high-peak and GLMS approximations is almost negligible in calculating the PBH abundance, and setting $\gamma\approx 1$ in the GLMS approximation (i.e., considering only the zeroth and first spectral moments) is reasonable and safe. This conclusion is also valid for the PBH mass windows at $10^{-17}M_\odot$ and $30M_\odot$.

\begin{figure}[htb]
\centering
\includegraphics[width=0.65\linewidth]{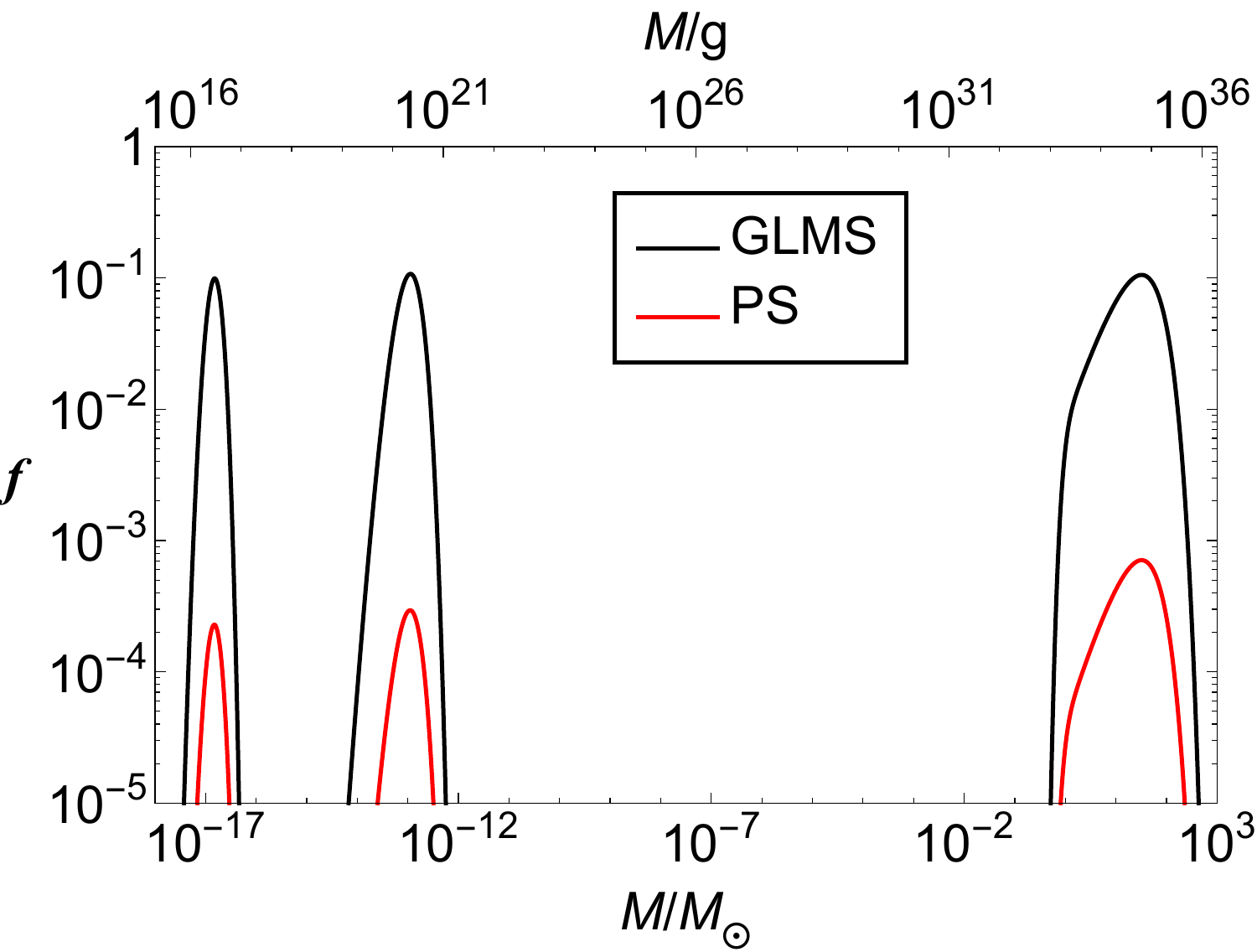}
\caption{The PBH abundances $f(M)$ obtained from the GLMS approximation (black lines) and the PS theory (red lines), with the PBH masses in the three typical mass windows at $10^{-17}M_\odot$, $10^{-13}M_\odot$, and $30M_\odot$, respectively. The $f(M)$ from the GLMS approximation are set to be $0.1$, and the $f(M)$ from the PS theory are only $2.30547\times 10^{-4}$, $2.73895\times 10^{-4}$, and $6.71296\times 10^{-4}$, respectively, two or three orders of magnitude lower. Therefore, the PS theory systematically underestimates the PBH abundance than peak theory.} \label{fig:compare}
\end{figure}

Furthermore, we compare the PBH abundances obtained from the GLMS approximation and the PS theory, with the PBH masses in the three typical mass windows. In Fig. \ref{fig:compare}, the PBH abundances obtained from the GLMS approximation are set to be $0.1$, and we observe that the PBH abundances obtained from the PS theory are merely $10^{-4}\sim 10^{-3}$, two or three orders of magnitude lower, indicating that the PS theory significantly underestimates the PBH abundance. This can be understood from Eq. (\ref{ln}). At the leading order of $\nu_{\rm c}$, we have
\begin{align}
\f{\beta_{\rm PS}}{\beta_{\rm GLMS}}\approx \f{1}{(Q\nu_{\rm c})^3}, \n
\end{align}
with $\nu_{\rm c}\gg 1$ taken into account. The numerical results of ${\beta_{\rm PS}}/{\beta_{\rm GLMS}}$ and the specific values of $\nu_{\rm c}$, $Q$, and $1/(Q\nu_{\rm c})^3$ are summarized in Tab. \ref{tab:3}, and we find that the theoretical values of $1/(Q\nu_{\rm c})^3$ agree with the numerical ratios of ${\beta_{\rm PS}}/{\beta_{\rm GLMS}}$ very well.

\begin{table}[htb]
\renewcommand\arraystretch{1.3}
\centering
\begin{tabular}{m{1.2cm}<{\centering} |m{1.7cm}<{\centering} |m{1.2cm}<{\centering} |m{1.4cm}<{\centering}| m{1.8cm}<{\centering}}
\hline\hline
$M/M_\odot$ & ${\beta_{\rm PS}}/{\beta_{\rm GLMS}}$ & $\nu_{\rm c}$ & $Q$ & $1/(Q\nu_{\rm c})^3$ \\
  \hline
$10^{-17}$  & 0.00230547 & 9.33184 & 0.814301 & 0.00227899 \\
  \hline
$10^{-13}$  & 0.00273195 & 8.82676 & 0.813913 & 0.00269689 \\
  \hline
$30$        & 0.00671296 & 6.57976 & 0.811961 & 0.00655791 \\
\hline\hline
\end{tabular}
\caption{The numerical ratios of the PBH mass fraction ${\beta_{\rm PS}}/{\beta_{\rm GLMS}}$ obtained from the GLMS approximation and the PS theory and the theoretical values of $1/(Q\nu_{\rm c})^3$ at the leading order of $\nu_{\rm c}$. As $\nu_{\rm c}\sim{\cal O}(10)\gg 1$, the theoretical results agree with the numerical ratios very well.} \label{tab:3}
\end{table}

Altogether, the PS theory has a systematic bias from the high-peak and GLMS approximations. Hence, the agreements of the PBH abundances obtained in Refs. \cite{Green:2004wb, Young:2014ana} apply only to the power spectra in the power-law form or with a running spectral index, but are no longer valid for the power spectrum with a peak on small scales due to the USR inflation. Moreover, it was also pointed out that the underestimation of the PBH abundance from the PS theory will be further enlarged, if the extended mass function is taken into account \cite{Wu:2020ilx}.

\section{Conclusion} \label{sec:con}

The research on PBHs is an interesting combination of black hole physics and cosmology, greatly promoted by the discovery of the merging GWs from binary black holes in recent years. One of the fundamental motivations to study PBHs is to seek an effective candidate for DM. Currently, compared with the WIMP-like or axion-like DM candidates, the experimental constraints on PBHs are still rather loose. Therefore, the aim of our work is to phenomenologically study the PBH abundance by considering the perturbations in the inflaton potential. We systematically calculate the power spectrum $\cP_\cR$ and the PBH abundance $f$ via the GLMS approximation of peak theory, with the high-peak approximation and the PS theory also carefully investigated for comparison. Our basic conclusions can be drawn as follows.

(1) The perturbations in the inflaton potential can significantly decelerate the local velocity of the inflaton, driving inflation into the USR stage, during which the usual SR approximations are invalidated, and the primordial curvature perturbation $\cR$ still increases outside the horizon. By this means, the power spectrum $\cP_\cR$ is remarkably enhanced on small scales, producing the PBHs with desirable masses and abundances.

(2) The specific form of the perturbation is constructed to be antisymmetric in the USR region, with three model parameters $A$, $\phi_0$, and $\sg$, so that the perturbation can be smoothly imposed on the background inflaton potential on both sides of the USR region. Thus, a perfect plateau can be created, and the USR stage can be maintained for a sufficiently long time. In this way, the adjustments of model parameters become much easier, and the fine-tuning problem frequently encountered in the USR inflation can be greatly relieved.

(3) In the case of one perturbation, we can achieve the PBHs with $f\sim 0.1$ in the three typical mass windows at $10^{-17}M_\odot$, $10^{-13}M_\odot$, and $30M_\odot$, respectively. With $M$ increasing, $\phi_0$ increases, as the large-mass PBH exits the horizon earlier. At the same time, $\sg$ decreases, as the inflaton rolls down more slowly in the early era, so the duration of the USR stage must be shortened; otherwise, $\cP_\cR$ would be too high.

(4) In the case of two perturbations, we can also achieve the PBHs of different masses both with $f\sim 0.1$ in two of the three typical mass windows. The distance between the two perturbations increases with that between the two mass windows, and the plateaus become more inclined when the two perturbations are close, in order to avoid their interference to a greater extent. Moreover, due to this interference, there is a turning point in the growing stage of $\cP_\cR$ on the small scale.

(5) The PBH abundances obtained from the high-peak and GLMS approximations are almost the same, because we always have $\nu_{\rm c}\gg 1$ and $\gamma\approx 1$. However, the PBH abundance obtained from the PS theory is two or three orders of magnitude lower. This underestimation is mainly due to the fact that the PS theory is based on a non-physical dimensional reduction of the general peak theory, so there is systematic bias between it and the high-peak or GLMS approximation, especially in the USR inflation. In general, peak theory is based on a sounder theoretical footing, and the PS theory should be regarded as its limit case, so the relevant results must be treated with great caution.

Last, we should mention that the PBH abundance is also affected by many other ingredients, among which the most influential is the threshold of density contrast $\dt_{\rm c}$ (or $\nu_{\rm c}$). From Eqs. (\ref{hp})--(\ref{betaPS}), we observe that the PBH mass fraction is exponentially sensitive to $\nu_{\rm c}$. The specific value of the threshold for the PBH formation is a long-debated issue in the literature \cite{Niemeyer:1999ak, Musco:2004ak, Musco:2008hv, Musco:2012au, 414, Nakama:2013ica, Musco:2018rwt, Escriva:2019nsa, Escriva:2019phb, Escriva:2020tak, Musco:2020jjb} and is affected by a number of aspects, such as the profile of density contrast, the equation of state of the cosmic medium, and the primordial non-Gaussianity. Therefore, in this paper, we follow the most frequently used value $\dt_{\rm c}=0.414$ \cite{414}. This is also the reason that we have not compared the PBH abundance with all currently available experimental constraints, because the focus of this work is not the elaborate adjustments of model parameters that can be equivalently replaced by a small change of the threshold, but is the antisymmetric construction of the perturbation in the inflaton potential and the comparison of peak and PS theories in calculating the PBH abundance.

\acknowledgments

We are very grateful to Zhao-Hui Chen, Florian K\"{u}hnel, Jing Liu, Dominik Schwarz, and Ji-Xiang Zhao for fruitful discussions. This work is supported by the Fundamental Research Funds for the Central Universities of China (No. N182410001).

%\paragraph{Note added.} This is also a good position for notes added after the paper has been written.

\end{document}